\documentclass[10pt,letterpaper,twoside,final,twocolumn]{IEEEtran}
\usepackage[english]{babel}
\usepackage{setspace}
\usepackage{fixltx2e}
\usepackage{graphicx}
\usepackage{subfig}
\usepackage{algorithm}
\usepackage{algorithmic}
\usepackage{epsfig}
\usepackage{epstopdf}
\usepackage[cmex10]{amsmath}
\usepackage{cite}
\usepackage{amssymb}
\usepackage{amsthm}
\usepackage{xcolor}
\usepackage{multirow}
\usepackage{array}
\usepackage{longtable}
\definecolor{mygreen}{rgb}{0,0.4,0}

\newcommand{\abs}[1]{\left| #1 \right|} 

\let\baraccent=\= 
\renewcommand{\=}[1]{\stackrel{#1}{=}} 

\usepackage[xcolor]{changebar}
\usepackage{xcolor}
\usepackage{tabularx}

\usepackage{epstopdf}
\DeclareGraphicsExtensions{.eps,.pdf,.png,.jpg}

\usepackage{mathtools}

\newtheorem{prop}{Proposition}

\theoremstyle{definition}
\newtheorem{dfn}{Definition}
\theoremstyle{definition}
\newtheorem{rel}{Relaxation}

\PassOptionsToPackage{hyphens}{url}\usepackage[breaklinks=true, colorlinks=true, citecolor=blue, bookmarksopen=true]{hyperref}

\nochangebars
\begin{document}
\title{
{A Quantum-Search-Aided Dynamic Programming Framework for Pareto Optimal Routing in Wireless Multihop Networks}}
\author{
Dimitrios~Alanis,~\IEEEmembership{Student~Member,~IEEE,}
Panagiotis~Botsinis,~\IEEEmembership{Member,~IEEE,} 
Zunaira~Babar,
Hung Viet Nguyen,~\IEEEmembership{Member,~IEEE,}
Daryus Chandra,~\IEEEmembership{Student~Member,~IEEE,} 
Soon~Xin~Ng,~\IEEEmembership{Senior~Member,~IEEE,}  and~Lajos~Hanzo,~\IEEEmembership{Fellow,~IEEE}\vspace*{-0.5cm}%
\thanks{The authors are with the School of Electronics and Computer Science, University of Southampton, Southampton, SO17 1BJ, UK (email: \{da1d16,~pb1y14,~zb2g10,~hvn08r,~dc2n14,~sxn,~lh\}@ecs.soton.ac.uk).}
\thanks{The financial support of the EPSRC under the grant EP/L018659/1, that of the European Research Council, Advanced Fellow Grant and that of the Royal Society's Wolfson Research Merit Award is gratefully acknowledged. Additionally, the authors acknowledge the use of the IRIDIS High Performance Computing Facility, and associated support services at the University of Southampton, in the completion of this work. The research data of this paper can be found at \url{http://doi.org/10.5258/SOTON/D0402}.}
}
\markboth{IEEE Transactions on Communications 2018}{D.~Alanis \emph{et al.}: A Quantum-aided Dynamic Programming Framework for Pareto Optimal Routing in WMHNs}
\maketitle
\begin{abstract}
Wireless Multihop Networks~(WMHNs) have to strike a trade-off among diverse and often conflicting Quality-of-Service~(QoS) requirements. The resultant solutions may be included by the Pareto Front under the concept of Pareto Optimality. However, the problem of finding all the Pareto-optimal routes in WMHNs is classified as NP-hard, since the number of legitimate routes increases exponentially, as the nodes proliferate. Quantum Computing offers an attractive framework of rendering the Pareto-optimal routing problem tractable. In this context, a pair of quantum-assisted algorithms have been proposed, namely the Non-Dominated Quantum Optimization~(NDQO) and the Non-Dominated Quantum Iterative Optimization~(NDQIO). However, their complexity is proportional to $\sqrt{N}$, where $N$ corresponds to the total number of legitimate routes, thus still failing to find the solutions in ``polynomial time''. As a remedy, we devise a dynamic programming framework and propose the so-called Evolutionary Quantum Pareto Optimization~(EQPO) algorithm. We analytically characterize the complexity imposed by the EQPO algorithm and demonstrate that it succeeds in solving the Pareto-optimal routing problem in polynomial time. Finally, we demonstrate by simulations that the EQPO algorithm achieves a complexity reduction, which is at least an order of magnitude, when compared to its predecessors, albeit at the cost of a modest heuristic accuracy reduction.
\end{abstract}

\section*{List of Acronyms}
\begin{flushleft} 
\noindent \begin{tabular}[h]{p{0.2\linewidth} p{0.72\linewidth}}
BBHT-QSA & Boyer-Brassard-H{\o}yer-Tapp Quantum Search Algorithm\\
BF & Brute Force\\
BER & Bit Error Ratio\\
CDP & Classical Dynamic Programming\\
CF(E) & Cost Function (Evaluation)\\
DHA & Durr-H{\o}yer Algorithm\\
DN & Destination Node\\
EQPO & Evolutionary Quantum Pareto Optimization\\
HP & Hardware Parallelism\\
MODQO & Multi-Objective Decomposition Quantum Optimization\\
\end{tabular}
\end{flushleft}

\begin{flushleft} 
\begin{tabular}[h]{p{0.2\linewidth} p{0.72\linewidth}}
MO-ACO & Multi-Objective Ant Colony Optimization\\
NDQO & Non-Dominated Quantum Optimization\\
(P-)NDQIO & (Preinitialized) Non-Dominated Quantum Iterative Optimization\\
NSGA-II & Non-dominated Sorting Genetic Algorithm II\\
OF & Objective Function\\
OPF-SR & Optimal Pareto Front Self-Repair \\
(O)PF & (Optimal) Pareto Front\\
QoS & Quality-of-Service\\
QP & Quantum Parallelism\\
RN & Relay Node\\
SN & Source Node\\
UV & Utility Vector\\
UF & Utility Function \\
WMHN & Wireless MultiHop Network
\end{tabular}
\end{flushleft} 

\section{Introduction}
\IEEEPARstart{T}{he} concept of \emph{Wireless Multihop Networks}~(WMHN) \cite{jiao2016backpressure} enables the communication of remote nodes by forwarding the transmitted packets through a cloud of mobile relays. Naturally, the specific choice of the relays plays a significant role in the performance of WMHNs \cite{alawieh2009mwh}, thus bringing their routing optimization in the limelight. Explicitly, optimal routing relies on a fragile balance of diverse and often conflicting \emph{Quality-of-Service}~(QoS) requirements \cite{chen2011fundamental}, such as the route's overall \emph{Bit-Error-Ratio}~(BER) or \emph{Packet Loss Ratio}~(PLR), its total power consumption, its end-to-end delay, the route's achievable rate, the entire system's sum-rate and its ``lifetime'' \cite{yetgin2015network}.

For the sake of taking into account multiple QoS requirements, several studies consider single-component \emph{Objective Functions}~(OF) as their optimization objectives. In this context, the metric of {Network Lifetime}~(NL) \cite{Abdulla:HYMN,yetgin2015network} has been utilized, which involves the routes' power consumption in conjunction with the nodes' battery levels. Additionally, the so-called \emph{Network Utility}~(NU) \cite{tan2015utility} also constitutes a meritorious single-component optimization OF. Apart  from the aforementioned QoS requirements, NU also takes into account the routes' achievable rate~\cite{shi2008cross}. In conjunction with the construction of aggregate functions, the authors of \cite{caleffi2012opera,banirazi2014heat} also incorporate QoS as constraints, thus providing a more holistic view of the routing problem. In this context, Banirazi \emph{et al.}~\cite{banirazi2014heat} optimized an aggregate function of the Dirichlet routing cost as well as the average network delay at specific operating points that maximize the network throughput.

The beneficial properties of \emph{dynamic programming} \cite{dasgupta2006algorithms} have been exploited for the sake of identifying the optimal routes, while relying on single-component aggregate functions. In this context, Dijkstra's algorithm~\cite{ramirez2013optimal,zuo2014cross,luo2014green} has been employed, since it is capable of approaching the optimal routes at the cost of imposing a complexity on the order of $O(E^3)$, where $E$ corresponds to the number of edges in the network's graph. Additionally, the appropriately modified Viterbi decoding algorithm \cite{you2012anear,wang2013dynamic} has also been utilized for solving single-component routing optimization problems, where the route exploration process can be viewed as a \emph{trellis graph} and thus the routing problem is transformed into a decoding problem. Explicitly, this transformation is reminiscent of the famous \emph{Bellman-Ford} algorithm~\cite{yao2016secure}.

The aforementioned approaches fail to identify the potential discrepancies among the QoS requirements, but they can be unified by the concept of \emph{Pareto Optimality}~\cite{deb2005mo}. However, the search-space of multi-component optimization is inevitably expanded due to combining the single-component OFs. Furthermore, the complexity is proportional to $O(N^2)$, where $N$ corresponds to the total number of eligible routes. Additionally, since $N$ increases exponentially as the relay nodes proliferate~\cite{Yetgin:NSGA_2}, the Pareto-optimal routing problem is classified as \emph{Non-deterministic Polynomial hard}~(NP-hard) \cite{alanis2014ndqo}. This escalating complexity can be partially mitigated by identifying a single Pareto-optimal solution. For instance, Gurakan \emph{et al.} \cite{gurakan2016optimal} conceived an optimal iterative routing scheme for identifying a single Pareto-optimal solution in terms of the sum rate and the energy consumption of wireless energy-transfer-enabled networks. However, in our application we are primarily interested in identifying the entire set of Pareto-optimal solution, since it provides fruitful insights into the underlying trade-offs \cite{deb2005mo}. In this context, multi-objective evolutionary algorithms \cite{Yetgin:NSGA_2,Camelo:NSGA, Martins:DCCP} have been employed for addressing the escalating complexity. In particular, Yetgin \emph{et al.} \cite{Yetgin:NSGA_2} used both the \emph{Non-dominated Sorting Genetic Algorithm II}~(NSGA-II) and the \emph{Multi-Objective Differential Evolution Algorithm}~(MODE) for optimizing the transmission routes in terms of their end-to-end delay and power dissipation. While considering a similar context, Camelo \emph{et al.} \cite{Camelo:NSGA} invoked the NSGA-II for optimizing the same QoS requirements for both the ubiquitous \emph{Voice over Internet Protocol}~(VoIP) and for file transfer. Additionally, the so-called \emph{Multi-Objective Ant Colony Optimization}~(MO-ACO) algorithm \cite{lopez2012theauto} has been employed in \cite{alanis2014ndqo} for the sake of addressing the multi-objective routing problem in WMHNs.

Quantum computing provides a powerful framework \cite{grover1996fast, PROP:PROP493, durr1996quantum} for the sake of rendering Pareto-optimal routing problems tractable by exploiting the so-called \emph{Quantum Parallelism}~(QP) \cite{nielsen2010quantum}. Explicitly, in \cite{wang2016quantum}  \emph{Quantum Annealing}~\cite{wang2016differential}, has been invoked for the sake of optimizing the activation of the wireless links in wireless networks, while maintaining  the maximum throughput and minimum interference as well as providing a substantial complexity reduction w.r.t. its classical counterpart, namely simulated annealing. In terms of Pareto optimal routing using \emph{universal quantum computing}~\cite{nielsen2010quantum}, the so-called \emph{Non-Dominated Quantum Optimization}~(NDQO) algorithm proposed in \cite{alanis2014ndqo} succeeded in identifying the entire set of Pareto-optimal routes at the expense of a complexity, which is on the order of $O(N\sqrt{N})$, relying on QP. As an improvement, the so-called \emph{Non-Dominated Quantum Iterative Optimization}~(NDQIO) algorithm was proposed in \cite{alanis2015ndqio}. Explicitly, the NDQIO algorithm is also capable of identifying the entire set of Pareto-optimal routes, while imposing a parallel complexity and a sequential complexity defined\footnote{We define the \emph{parallel complexity} as the complexity imposed while taking into account the degree of parallelism. By contrast, the sequential complexity does not consider any kind of parallelism. In \cite{alanis2015ndqio}, they are referred to as \emph{normalized execution time} and \emph{normalized power consumption}, respectively.} in \cite{alanis2015ndqio}, which is on the order of $O(N_\text{OPF}\sqrt{N})$ and $O(N^2_\text{OPF}\sqrt{N})$, respectively, by relying on the beneficial synergy between QP and \emph{Hardware Parallelism}~(HP). Note that $N_\text{OPF}$ corresponds to the number of Pareto-optimal routes.

Despite the substantial complexity reduction offered both by the NDQO and the NDQIO algorithms, the multi-objective problem still remains intractable, when the network comprises an excessively high number of nodes due to the escalating complexity. Explicitly, Zalka~\cite{zalka1999grover} has demonstrated that the complexity order of $O(\sqrt{N})$ is the minimum possible, as long as the database values are uncorrelated. By contrast, when the formation of the Pareto-optimal route-combinations becomes correlated owing to socially-aware networking \cite{alanis2016modqo}, a further complexity reduction can be achieved. Based on this specific observation, we will design a novel algorithm, namely the \emph{Evolutionary Quantum Pareto Optimization}~(EQPO), in order to exploit the correlations exhibited by the individual Pareto-optimal routes by appropriately constructing trellis graphs that guide the search process in the same fashion as in Viterbi decoding. Furthermore, we will also exploit the synergies between QP and HP for the sake of achieving an additional complexity reduction by considering as low a fraction of the database entries as possible, while still guaranteeing a near-full-search-based performance. 

Our contributions are summarized as follows:
\emph{
\begin{enumerate}
\item[\emph{1)}] In Section~\ref{sec:mordp}, we develop a novel  multi-objective dynamic programming framework for generating potentially Pareto-optimal routes relying on the correlations of the specific links constituting the Pareto-optimal routes, hence substantially reducing the total number of routes considered. Explicitly, this framework is a multi-objective extension of the popular single-objective Bellman-Ford algorithm.
\item[\emph{2)}] In Section~\ref{sec:eqpo}, we propose a novel quantum-assisted algorithm, namely the \emph{Evolutionary Quantum Pareto Optimization} algorithm, which jointly exploits our novel dynamic programming framework as well as the synergies between the QP and the HP for the sake of solving the multi-objective routing problem of WMHNs. 
\item[\emph{3)}] In Section~\ref{sec:accVScomp}, we also characterize the performance versus complexity of the EQPO algorithm and demonstrate that it achieves both a parallel and a sequential complexity reduction of at least an order of magnitude for a 9-node WMHN, when compared to that of the NDQIO algorithm. 
\end{enumerate}
}

The rest of this paper is organized as follows. In Section~\ref{sec:netSec}, we will briefly discuss the specifics of the network model considered in our case study. In Section~\ref{sec:mordp}, we will present a dynamic programming framework, which is optimal in terms of its heuristic accuracy. In Section~\ref{sec:eqpo}, we will relax the optimal framework of Section~\ref{sec:netSec} for the sake of striking a better accuracy versus complexity  trade-off with the aid of our EQPO algorithm. Subsequently, in Section~\ref{subsec:complexity} we will analytically characterize the EQPO algorithm's complexity and in Section~\ref{subsec:accuracy} we will evaluate its performance.

\section{Network Specifications}\label{sec:netSec}
\begin{figure}[htb]
\centering
\includegraphics[scale=0.8]{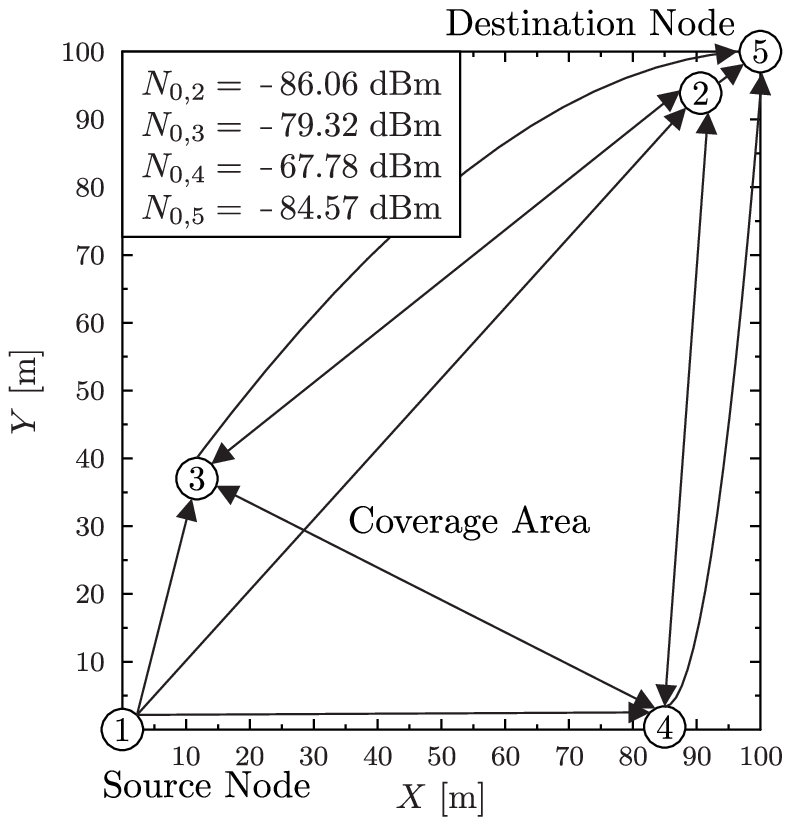}
\caption{Exemplified WMHN topology associated with 5 nodes. The presence of a cluster-head in possession of a quantum computer is considered at the DN side as in \cite{alanis2014ndqo} and in \cite{alanis2015ndqio}. The interference levels experienced by each node are presented in the legend.\label{fig:network-topology}}
\end{figure}

In the context of this treatise, the model of the networks considered both in \cite{alanis2014ndqo} and in \cite{alanis2015ndqio} has been adopted. To elaborate further, the WMHN considered is a fully connected network and it consists of a single \emph{Source Node}~(SN), a single \emph{Destination Node}~(DN) and a cloud of \emph{Relay Nodes}~(RN). The SN and the DN are located in the opposite corners of a (100$\times$100) m$^2$ square-block area, which is the WMHN coverage area considered. By contrast, the RNs are considered to be roaming within the coverage area  having random locations, which obey the uniform distribution within the WMHN coverage area. A WMHN topology is exemplified in Fig.~\ref{fig:network-topology} for a WMNH consisting of 5 nodes in total. Additionally, a cluster-head equipped with a quantum computer, which is responsible for collecting all the required WMHN information, such as the nodes' geolocations and their interference levels, is considered to be present at the DN side. Therefore, we should point out that this treatise is focused on a centralized protocol.

Based on the network information gathered, the WMHN cluster-head has to identify the optimal routes emerging from the SN to the DN based on certain \emph{Utility Functions} (UF). Similar to \cite{alanis2014ndqo} and \cite{alanis2015ndqio}, we have jointly taken into account the route's overall delay, its overall  power consumption and its overall \emph{Bit Error Ratio} (BER). Before delving into the UFs, let us define a legitimate route of our WMHN consisting of $N_\text{nodes}$ nodes, as $x_r = [\text{SN},\dots,\text{DN}]$, which contains each RN only once for the sake of limiting the total number $N$ of legitimate routes, while at the same time avoiding routes associated with excessive power consumption and delay. Note that we have associated the SN and the DN with the node indices 1 and $N_\text{nodes}$, respectively, in the context of this treatise. Additionally, these legitimate routes are mapped to a specific index $x$ under lexicographic ordering using \emph{Lehmer Encoding}\footnote{Lehmer Encoding maps a specific permutation  to an index in the \emph{factoradic basis} \cite{lehmer1960teaching}.} \cite{lehmer1960teaching}. The route's overall delay $D(x)$ is considered as one of our UFs, which is quantified in terms of the number of hops established by the route. This is formally formulated as follows:
\begin{equation}\label{eq:delay}
D(x) = \abs{x_r}-1,
\end{equation}
where the operator $\abs{\cdot}$ corresponds to the number of nodes along the route $x_r$ including the SN and DN. Moving on to the $x$-th route's overall power consumption $P(x)$, it is proportional to the sum of path-losses incurred by each of the individual links constituting the route. Explicitly, the path-loss $L_\text{dB}(i,j)$ quantified in dB for a single link between the $i$-th and the $j$-th nodes is equal to \cite{alanis2015ndqio}:
\begin{equation}
L_\text{dB}(i,j) = P_{Tx,ij} - P_{Rx,ij}=  10\alpha\log_{10}\left(\frac{4 \pi d_{i,j}}{\lambda_c}\right),
\end{equation}
where $\alpha$ corresponds to the path-loss exponent, $d_{i,j}$ is the distance between the two nodes and $\lambda_c$ denotes the carrier's wavelength. In our case-study we have set $\alpha=3$ and $\lambda_c\simeq 0.125$ m corresponding to a frequency of $f_c=2.4$ GHz. Consequently, the second UF is formulated as follows:
\begin{equation}\label{eq:power}
L(x) = \sum\limits_{i=1}^{\abs{x_r}-1}10^{L_\text{dB}(x_{r}^{(i)},x_{r}^{(i+1)})/10}.
\end{equation}

Moving on to the final UF, namely the BER, let us first elaborate on the interference levels experienced by the nodes. In our specific scenario, there is only a single pair of source and destination nodes, resulting in a single route being active. Additionally, we have assumed that the WMHN has a sufficient number of orthogonal spreading codes and sub-carriers for the sake of efficiently separating the routes as in \cite{alanis2016modqo}. In this context, there is no interference stemming from the WMHN itself; however, we have assumed that a sufficiently high number of users access the channel, hence the resultant interference can be treated as \emph{Additive White Gaussian Noise} (AWGN), owing to the \emph{Central Limit Theorem} (CLT) \cite{steele1999mobile}. Therefore, the interference is modeled by a random Gaussian process, with its mean set to -90 dBm and its standard deviation set to 10 dB, while the transmission power is set to $P_{Tx}=20$~dBm. Additionally, the nodes transmit their messages using the uncoded QPSK scheme \cite{hanzo2004single} over uncorrelated Rayleigh fading channels and utilize \emph{Decode-and-Forward} relaying \cite{yang2015isthe} for forwarding the respective messages. Based on these assumptions, we can readily use the closed-form BER performance of the adopted scheme versus the received \emph{Signal-to-Noise Ratio} (SNR), while the overall route's BER $P_e(x)$ can be calculated using the following recursive formula \cite{alanis2014ndqo}:
\begin{equation}\label{eq:ber_rec}
P_{e,tot}=P_{e,1}+P_{e,2}-2P_{e,1}P_{e,2},
\end{equation}
which corresponds to the output BER $P_{e,tot}$ of a two-stage \emph{Binary Symmetric Channel} (BSC) \cite{alanis2014ndqo}, where $P_{e,1}$ and $P_{e,2}$ represent the BER associated with the first and the second stage, respectively. 

Having described the UFs considered, let us now proceed by defining our optimization problem. Explicitly, we will jointly consider the UFs in the form of a \emph{Utility Vector}~(UV) $\mathbf{f}(x)$, which is defined as follows:
\begin{equation}\label{eq:uv}
\mathbf{f}(x) = \left[ P_e(x), L(x), D(x) \right],
\end{equation}
where $D(x)$ and $L(x)$ correspond to the $x$-th route's delay and power consumption defined in Eqs.~(\ref{eq:delay}) and (\ref{eq:power}), while $P_e(x)$ denotes the $x$-th route's end-to-end BER, which is recursively evaluated using Eq.~(\ref{eq:ber_rec}). Explicitly, we opt for jointly minimizing the entire set of UFs considered by the UV of Eq.~(\ref{eq:uv}). Therefore, for the evaluation of the fitness of the UVs we will utilize the concept of \emph{Pareto Optimality}\footnote{The readers should refer to \cite{alanis2016modqo} for a more detailed tutorial on Pareto optimality.}~\cite{deb2005mo}, which is encapsulated by Definitions~\ref{dfn:pd} and \ref{dfn:po}.
\begin{dfn}
\label{dfn:pd} {\bf Pareto Dominance} ~\cite{deb2005mo}: A particular route $x_i$ associated with the UV $\mathbf{f}(x_i) = [f_1(x_i),\dots,$ $f_K(x_i)]$, where $K$ is the number of the UFs considered, is said to strongly dominate another route  $x_j$ associated with the UV $\mathbf{f}(x_j) = [f_1(x_j),\dots, f_K(x_j)]$, denoted by $\mathbf{f}(x_i)\succ\mathbf{f}(x_j)$, iff we have $f_k(x_i)<f_k(x_j)$, $\forall{k}\in\{1,\dots,K\}$. Equivalently, the route $x_i$ is said to weakly dominate another route  $x_j$, denoted by $\mathbf{f}(x_i)\succeq\mathbf{f}(x_j)$, iff we have $f_k(x_i)\leq f_k(x_j)$, $\forall{k}\in\{1,\dots,K\}$ and $\exists k^\prime\in\{1,\dots,K\}$, so that we have $f_{k^\prime}(x_i)<f_{k^\prime}(x_j)$.
\end{dfn}
\begin{dfn}
\label{dfn:po} {\bf Pareto Optimality}~\cite{deb2005mo}: A particular route $x_i$ associated with the UV $\mathbf{f}(x_1)$ is Pareto-optimal, iff there is no route that dominates $x_i$, i.e. we have $\nexists{x_j}$ so that  $\mathbf{f}(x_j)\succ\mathbf{f}(x_i)$ is satisfied. Equivalently, the route $x_i$ is strongly Pareto-optimal  iff there is no route that weakly dominates $x_i$, i.e. we have $\nexists{x_j}$, so that $\mathbf{f}(x_i)\succeq\mathbf{f}(x_j)$ is satisfied.
\end{dfn}
Explicitly, Definition~\ref{dfn:pd} provides us with the criterion for evaluating the fitness of a specific route with respect to another reference route, while Definition~\ref{dfn:po} outlines the condition of the specific route's optimality. Based on the number of routes dominating a specific route, it is possible to group the routes into the so-called \emph{Pareto Fronts}~(PF). Explicitly, the PF comprises the Pareto-optimal routes, which are dominated by no other routes according to Definition~\ref{dfn:po}, which is often referred to as the \emph{Optimal Pareto Front}~(OPF). 

In our application, our aim is to identify the entire set of weakly Pareto-optimal routes for the sake of gaining insight into the routing trade-offs associated with the UFs considered. Naturally, for the sake of identifying a specific route as Pareto-optimal we have to perform precisely $(N-1)$ Pareto-dominance comparisons, where $N$ corresponds to the total number of legitimate routes. Therefore, the complexity imposed by the exhaustive search aiming for identifying the entire set of routes  belonging to the OPF is on the order of $O(N^2)$. Explicitly, the total number $N$ of legitimate routes increases exponentially as the number $N_\text{nodes}$ of nodes increases \cite{alanis2014ndqo}, hence rendering the multi-objective routing problem as NP-hard. Thus sophisticated methods are required for finding all of the solutions.

Let us now proceed by elaborating on our novel dynamic framework designed for efficiently exploring the search space.

\section{Mutli-Objective Routing Dynamic Programming Framework}\label{sec:mordp}

Before delving into the analysis of our multi-objective dynamic programming framework, which is specifically tailored for our routing problem, we will express each of the UFs considered in the UV of Eq.~(\ref{eq:uv}) as a weighted sum of the specific UFs associated with the individual links comprised by a particular route. Explicitly, the power consumption has already been expressed in this form based on Eq.~(\ref{eq:power}). As for the delay, which we have defined as the number of hops, it may be redefined as follows:
\begin{equation}\label{eq:delay_convex}
D(x) = \sum\limits_{i=1}^{\abs{x_r}-1}\left(1-\delta_{x_r^{(i)},x_r^{(i+1)}}\right),
\end{equation}
where $\delta_{i,j}$ corresponds to the \emph{Kronecker delta} function \cite{abramowitz1966handbook}, while $x_r$ and $x$ represent the route and its associated index, respectively. As for the route's overall BER, the recursive formula of Eq.~(\ref{eq:ber_rec}) may be approximated as follows:
\begin{equation}\label{eq:ber_convex}
P_e(x) = \sum\limits_{i=1}^{\abs{x_r}-1}P_{e,x_r^{(i)},x_r^{(i+1)}}-\epsilon(x)\thickapprox \sum\limits_{i=1}^{\abs{x_r}-1}P_{e,x_r^{(i)},x_r^{(i+1)}},
\end{equation}
where $P_{e,k,l}$ represents the BER of the specific link established between the $k$-th and the $l$-th nodes, while $\epsilon(x)$ is the approximation error, which is on the order of:
\begin{equation}\label{eq:ber_approx_error}
\epsilon(x) = O\left( \sum\limits_{i=1}^{\abs{x_r}-1}\sum\limits_{\scriptsize\begin{array}{c}
j=1\\ j\neq i
\end{array}}^{\abs{x_r}-1}P_{e,x_r^{(j)},x_r^{(i+1)}}P_{e,x_r^{(j)},x_r^{(j+1)}}\right).
\end{equation}
Since the sum of the products of all the links' BER will be several orders of magnitude lower than their sum, the approximation error of Eq.~(\ref{eq:ber_convex}) may be deemed to be negligible. 

Having expressed the UFs considered as a weighted sum of the UFs associated with their links, we may now proceed by exploiting this specific property for the sake of achieving a further complexity reduction. In fact, it is possible to transform our composite multi-objective routing problem into a series of smaller subproblems, thus arriving at a dynamic programming structure. This transformation is performed with the aid of Definition~\ref{def:route_generation} in conjunction with Proposition~\ref{prop:route_optimality}. 

\begin{dfn}\label{def:route_generation}
A specific route $x=\{SN{\rightarrow}\bar{R}_i{\rightarrow}DN\}$ is said to generate another route $x_g^{(j)}$ by inserting the single RN $R_j$ node between the previous RN and the DN. Explicitly, the resultant route $x_g^{(j)}$ is $x^{(j)}_g=\{SN{\rightarrow}\bar{R}_i{\rightarrow}R_j{\rightarrow}DN\}$, $\forall j \in \{1,\dots,N_\text{nodes}-2\}$.
\end{dfn}

\begin{prop}\label{prop:route_optimality}
Let us consider a specific route $x=\{SN{\rightarrow}\bar{R}_i{\rightarrow}DN\}$ associated with the UV $\mathbf{f}(x)= [f_1(x),\dots,f_K(x)]$ and its sub-route  $x^\prime=\{SN{\rightarrow}\bar{R}_i\}$ associated with the UV $\mathbf{f}(x^\prime)= [f_1(x^\prime),\dots,f_K(x^\prime)]$. Let us assume furthermore that each component $f_k(x)$ of the UV associated with the route $x$ has a positive value and that it can be expressed as a sum of the respective UFs of its individual links $x_{i,i+1}$, i.e. we have:

\begin{equation}\label{eq:uf_prot}
f_k(x)=\sum\limits^{\abs{x}-1}_{i=1}f_k(x_{i,i+1}),
\end{equation}
with ${f}_k(x_{i,i+1})>0,~\forall~k,i,x:~k\in\{1,...,K\},~ i\in\{1,...,|x|-1\},~x\in S$, where $K$ and $S$ are the number of optimization objectives and the set of legitimate routes, respectively. The route $x$ cannot generate any Pareto-optimal routes using the rule of Definition~\ref{def:route_generation} if there is a route  $x_d=\{SN{\rightarrow}\bar{R}_j{\rightarrow}DN\}$ from the SN to the DN associated with $\bar{R}_j\neq \bar{R}_i$ that weakly dominates the sub-route $x^{\prime}$, i.e. if we have $\exists x_d\in S: \mathbf{f}(x_d)\succeq \mathbf{f}(x^\prime)$. The respective proof is presented in Appendix~\ref{app:proof}.
\end{prop}

Explicitly, Proposition~\ref{prop:route_optimality} guarantees that a specific route $x=\{SN{\rightarrow}\bar{R}_i{\rightarrow}DN\}$ comprised by the sub-route $x^\prime=\{SN{\rightarrow}\bar{R}_i\}$ cannot generate Pareto-optimal routes by  adding an intermediate RN to $x$ between its last RN and the DN, if the sub-route $x^\prime$ is weakly dominated by any of the legitimate routes. Explicitly, should its sub-route $x^\prime$ be sub-optimal, the respective route $x$ will be sub-optimal as well, since we have $\exists x_d\in S: \mathbf{f}(x_d)\succeq \mathbf{f}(x^\prime)\succ \mathbf{f}(x)$, based on Proposition~\ref{prop:route_optimality}. Note that the opposite of this statement does not apply, since there exist sub-optimal routes, whose sub-routes are indeed Pareto-optimal. 

\begin{figure*}[thb]
\begin{center}
\includegraphics[width=\linewidth]{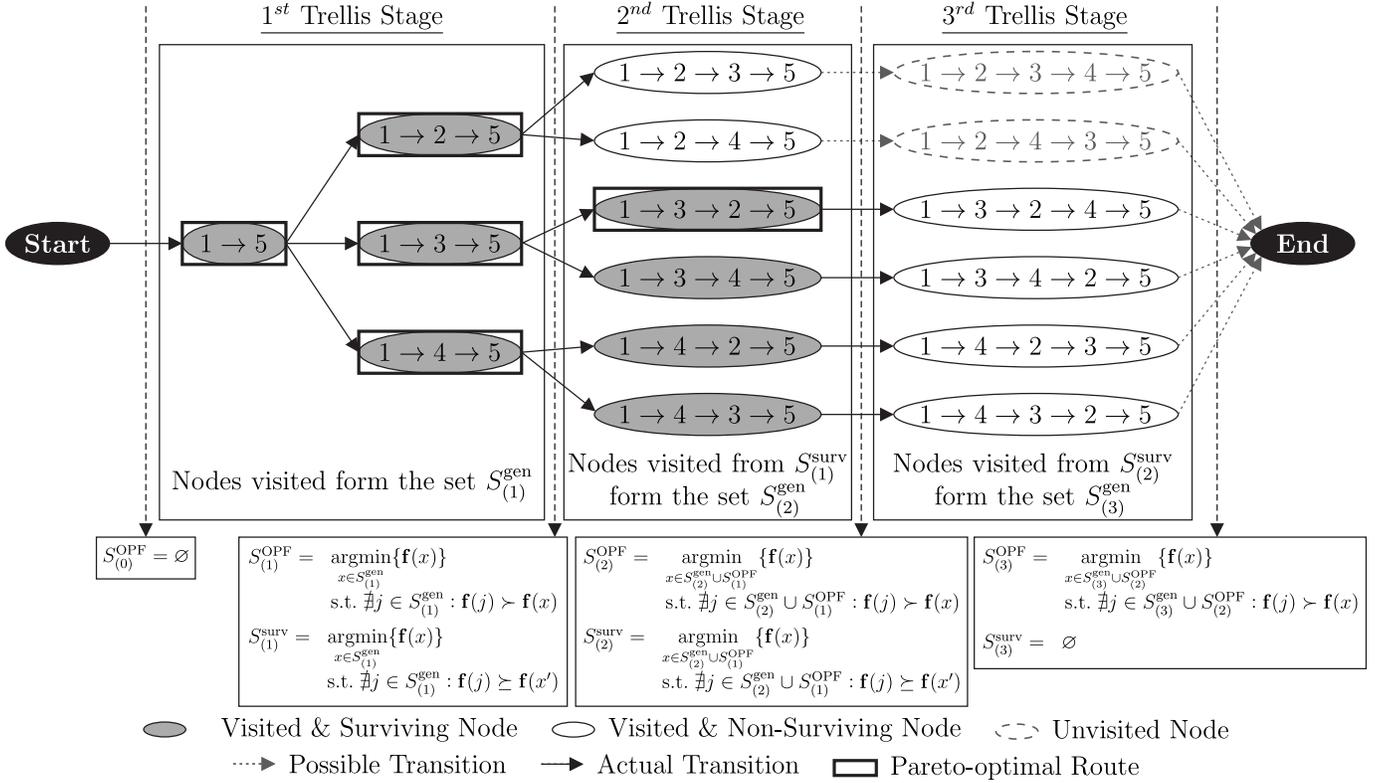}
\end{center}
\caption{Irregular trellis graph designed for guided search-space exploration for the 5-node WMHN of Fig.~\ref{fig:network-topology} using the optimal dynamic programming framework, encapsulated by Definition~\ref{def:route_generation} and Proposition~\ref{prop:route_optimality}. Note that the UVs of each route are presented in Table~\ref{tab:tut_uvs}.\label{fig:optimal-trellis}}
\end{figure*}\hfill\nopagebreak
\begin{table*}[h]
\begin{center}
\caption{Utility Vectors of the legitimate routes and of their respective sub-routes for the 5-node WMHN topology of Fig.~\ref{fig:network-topology}.\label{tab:tut_uvs}}
\begin{tabular}{c|c|c|c|c}
\hline
Route $x$ & Route UV & Sub-route UV & Optimal Route & Optimal Sub-route \\
\hline
$\{1~5\}$ & $[4.52~10^{-4},74.15,1]$ & $[\infty,\infty,\infty]$ & \checkmark & \checkmark\\
\hline
$\{1~2~5\}$ & $[2.52~10^{-4},73.10,2]$ & $[2.52~10^{-4},73.10,1]$ & \checkmark & \checkmark\\
$\{1~3~5\}$ & $[2.35~10^{-4},70.89,2]$ & $[3.13~10^{-5},57.30,1]$ & \checkmark & \checkmark \\
$\{1~4~5\}$ & $[1.43~10^{-2},71.76,2]$ & $[1.41~10^{-2},67.50,1]$ & \checkmark & \checkmark \\
\hline
$\{1~2~3~5\}$ & $[9.49~10^{-4},76.09,3]$ & $[7.45~10^{-4},74.61,2]$ & &\\
$\{1~2~4~5\}$ & $[1.91~10^{-2},75.72,3]$ & $[1.89~10^{-2},74.46,2]$ & &\\
$\{1~3~2~5\}$ & $[1.36~10^{-4},69.55,3]$ & $[1.36~10^{-4},69.54,2]$ & \checkmark & \checkmark\\
$\{1~3~4~5\}$ & $[1.29~10^{-2},71.74,3]$ & $[1.28~10^{-2},67.46,2]$ & & \checkmark \\
$\{1~4~2~5\}$ & $[1.42~10^{-2},71.19,3]$ & $[1.42~10^{-2},71.19,2]$ & & \checkmark \\
$\{1~4~3~5\}$ & $[1.46~10^{-2},73.50,3]$ & $[1.44~10^{-2},70.27,2]$ & & \checkmark \\
\hline
$\{1~2~3~4~5\}$ & $[1.36~10^{-2},76.36,4]$ & $[1.34~10^{-2},75.30,3]$ & &\\
$\{1~2~4~3~5\}$ & $[1.94~10^{-2},76.50,4]$ & $[1.92~10^{-2},75.18,3]$ & &\\
$\{1~3~2~4~5\}$ & $[1.90~10^{-2},74.13,4]$ & $[1.88~10^{-2},72.18,3]$ & &\\
$\{1~3~4~2~5\}$ & $[1.28~10^{-2},71.18,4]$ & $[1.28~10^{-2},71.17,3]$ & &\\
$\{1~4~2~3~5\}$ & $[1.49~10^{-2},75.23,4]$ & $[1.47~10^{-2},73.35,3]$ & &\\
$\{1~4~3~2~5\}$ & $[1.45~10^{-2},72.82,4]$ & $[1.45~10^{-2},72.81,3]$ & &\\
\hline
\end{tabular}
\end{center}
\end{table*} 

This specific property can be exploited for the sake of reducing the search-space size required for identifying the entire OPF. To elaborate further, we can devise an \emph{irregular trellis graph}~\cite{wenbo2015irrtrellis} for the sake of guiding the search space exploration, as portrayed in Fig.~\ref{fig:optimal-trellis} for the 5-node WMHN of Fig.~\ref{fig:network-topology}. Note however that this specific trellis graph is different from those used for channel coding in \cite{wenbo2015irrtrellis}, since  in the latter we only have as many legitimate paths as many legitimate symbols. By contrast, here all transitions represent legitimate routes in our scenario.  Additionally, we rely on Definition~\ref{def:route_generation} for the sake of determining the possible trellis-node transitions. For instance, observe in Fig.~\ref{fig:optimal-trellis} that a trellis-path emerging from the trellis-node associated with the generator route $\{1\rightarrow 2 \rightarrow 5\}$ is only capable of visiting the nodes associated with the routes $\{1\rightarrow 2 \rightarrow 3 \rightarrow 5\}$ and $\{1\rightarrow 2 \rightarrow 4 \rightarrow 5\}$, since a single RN is inserted before the DN into the generator route based on Definition~\ref{def:route_generation}. Moving on to the next trellis stages, during the $i$-th trellis stage the following three steps are carried out:
\subsubsection{Surviving Routes} The set $S^{\text{gen}}_{(i)}$ of generated routes are constructed based on the set $S^{\text{surv}}_{(i-1)}$ of surviving routes of the previous stage and relying on Definition~\ref{def:route_generation}.
\subsubsection{Pareto-Optimal Routes}The set $S^{\text{OPF}}_{(i)}$ of Pareto-optimal routes is identified based on the following optimization problem:
\begin{equation}\label{eq:sopf_i_optim}
\begin{array}{rl}
S^{\text{OPF}}_{(i)} =& \mathop{\text{argmin}}\limits_{x\in{S^{\text{gen}}_{(i)}\cup S^{\text{OPF}}_{(i-1)}}}\{\mathbf{f}(x)\},\\
{}& s.t.~ \nexists j\in{S^{\text{gen}}_{(i)}\cup S^{\text{OPF}}_{(i-1)}}: \mathbf{f}(j)\succ\mathbf{f}(x).
\end{array}
\end{equation}
Note that the optimization problem of Eq.~\eqref{eq:sopf_i_optim} considers the joint search space constituted by the all the routes $S^{\text{gen}}_{(i)}$ of the $i$-th trellis stage as well as by the Pareto-optimal routes $S^{\text{OPF}}_{(i-1)}$ of the previous stage. Using recursion, we can readily observe that the Pareto-optimal routes $S^{\text{OPF}}_{(i-1)}$ of the previous stage contain the Pareto-optimal routes across all stages up to the $(i-1)$-st stage. This property is beneficial for our dynamic programming framework, since it eliminates the need for backwards propagation, thus only requiring the employment of a feed-forward method for the identification of the entire OPF. 
\subsubsection{Surviving Routes}The set $S^{\text{surv}}_{(i)}$ of surviving routes is identified based on the following optimization problem:
\begin{equation}\label{eq:ssurv_i_optim}
\begin{array}{rl}
S^{\text{surv}}_{(i)} =& \mathop{\text{argmin}}\limits_{x\in{S^{\text{gen}}_{(i)}}}\{\mathbf{f}(x)\},\\
{}& s.t.~ \nexists j\in{S^{\text{gen}}_{(i)}\cup S^{\text{OPF}}_{(i-1)}}: \mathbf{f}(j)\succeq\mathbf{f}(x^\prime).
\end{array}
\end{equation}
where $x^\prime$ corresponds to the particular sub-route of $x$, having all the links of $x$, except for the last hop, as detailed in Proposition~\ref{prop:route_optimality}.

The optimization process proceeds to the next trellis stage as long as either there exist surviving routes, i.e. we have $S^{\text{surv}}_{(i)}\neq \varnothing$, or if the maximum affordable number of trellis stages - which is equal to the maximum number of hops of the legitimate routes - has not been exhausted. Otherwise, the optimization process terminates by exporting the hitherto identified OPF.

Let us now proceed by elaborating on the route exploration process using the 5-node WMHN example of Fig~\ref{fig:network-topology}. Its respective trellis is portrayed in Fig~\ref{fig:optimal-trellis}, while the routes' and their respective sub-route's UVs are shown in Table~\ref{tab:tut_uvs}. Initially, the optimization process considers the set $S^{\text{gen}}_{(1)}$ of routes, which is constituted by all the legitimate routes having a single and two hops, namely the routes  $\{1\rightarrow 5\}$, $\{1\rightarrow 2 \rightarrow 5\}$, $\{1\rightarrow 3 \rightarrow 5\}$ and $\{1\rightarrow 4 \rightarrow 5\}$, as portrayed in the $1^{st}$ trellis stage of Fig.~\ref{fig:optimal-trellis}. Based on Table~\ref{tab:tut_uvs}, all the routes considered are Pareto-optimal and thus the respective set is equal to $S^{\text{OPF}}_{(1)}=S^{\text{gen}}_{(1)}$. Subsequently, the set of surviving nodes is constructed. Explicitly, the direct route is not considered in this case, since its inclusion leads to the generation of routes, which have already been processed. Observe in Table~\ref{tab:tut_uvs} that all the routes constituted by 2 hops have Pareto optimal sub-routes and hence the set of surviving routes becomes $S^{\text{surv}}_{(1)} = \left[ \{1\rightarrow 2 \rightarrow 5\}, \{1\rightarrow 3 \rightarrow 5\},\{1\rightarrow 4 \rightarrow 5\}\right]$. 

After the identification of the set of surviving routes $S^{\text{surv}}_{(1)}$, the set $S^{\text{gen}}_{(2)}$ of routes generated in the 2$^{nd}$ trellis stage is created by including an appropriate RN right before the DN, as annotated with the aid of black arrows in Fig~\ref{fig:optimal-trellis}. Naturally, since all the routes constituted by two hops have been identified as being Pareto-optimal, the entire set of routes having three hops is visited by the trellis-paths in the 2$^{nd}$ trellis stage, as seen in Fig.~\ref{fig:optimal-trellis}. The set $S^{\text{OPF}}_{(1)}$ of Pareto-optimal routes of the $1^{st}$ trellis stage is then concatenated to the set $S^{\text{gen}}_{(2)}$ of the routes generated in the $2^{nd}$ trellis stage and the set $S^{\text{OPF}}_{(2)}$ of Pareto-optimal routes is identified. After this operation, the latter is set to $S^{\text{OPF}}_{(2)}=\left[ \{1\rightarrow 5\},\{1\rightarrow 2 \rightarrow 5\}, \{1\rightarrow 3 \rightarrow 5\},\{1\rightarrow 4 \rightarrow 5\}\right.$, $\left.\{1\rightarrow 3 \rightarrow 2 \rightarrow 5\}\right]$, hence including the route $\{1\rightarrow 3 \rightarrow 2 \rightarrow 5\}$ to the OPF, as denoted with the aid of the bold rectangle in Fig.~\ref{fig:optimal-trellis}. The surviving routes of the $2^{nd}$ trellis stage are then identified using the optimization problem of Eq.~\eqref{eq:ssurv_i_optim}. Explicitly, they constitute the set $S^{\text{surv}}_{(2)}=\{1\rightarrow 3 \rightarrow 2 \rightarrow 5\},\{1\rightarrow 3 \rightarrow 4 \rightarrow 5\},\{1\rightarrow 4 \rightarrow 2 \rightarrow 5\},\{1\rightarrow 4 \rightarrow 3 \rightarrow 5\}$, as it may be verified by Table~\ref{tab:tut_uvs} and denoted with the aid of the gray-filled nodes of Fig.~\ref{fig:optimal-trellis}. 

In the presence of surviving nodes, the optimization process proceeds with the final trellis stage; however, in this case the routes $\{1\rightarrow 2 \rightarrow 3 \rightarrow 4 \rightarrow 5\}$ and $\{1\rightarrow 2 \rightarrow 4  \rightarrow 3 \rightarrow 5\}$ are not considered, since their generators do not have Pareto-optimal sub-routes. This is portrayed in Fig.~\ref{fig:optimal-trellis} with the aid both of the gray dashed arrows and of the gray dashed nodes. Hence, the set $S^{\text{gen}}_{(3)}=\{1\rightarrow 3 \rightarrow 2 \rightarrow 4 \rightarrow 5\},\{1\rightarrow 3 \rightarrow 4 \rightarrow 2 \rightarrow 5\},\{1\rightarrow 4 \rightarrow 2 \rightarrow 4 \rightarrow 5\},\{1\rightarrow 4 \rightarrow 3 \rightarrow 2 \rightarrow 5\}$ is generated. The set $S^{\text{OPF}}_{(2)}$  is then concatenated to that of the routes generated in the final trellis stage and the final set $S^{\text{OPF}}_{(3)}$ of Pareto-optimal routes is identified. Explicitly, the latter is identical to the respective set of the $2^{nd}$ trellis stage, since none of the routes generated in the final stage is Pareto-optimal, as verified by Table~\ref{tab:tut_uvs}. Additionally, since we have reached the final stage, the set of surviving routes is not identified and the process exits by exporting the hitherto observed OPF.

In a nutshell, this route exploration process succeeds in transforming the multi-objective routing problem into a series of significantly less complex sub-problems, each corresponding to a single trellis stage, hence inheriting the structure of dynamic programming problems~\cite{dasgupta2006algorithms}. Note that the metric-accumulation, which is typical in dynamic programming problems, is constituted by the update of the Pareto-optimal routes. Note that this dynamic programming framework is optimal in terms of its efficacy in identifying the entire OPF, just like the exhaustive search method. Primarily, this is a benefit of Proposition~\ref{prop:route_optimality}, which excludes the routes that are incapable of generating Pareto-optimal routes during the next trellis stages.

\section{Evolutionary Quantum Pareto Optimization}\label{sec:eqpo} 
In Section~\ref{sec:mordp}, we introduced a novel dynamic programming framework for the sake of guiding the search process in identifying the Pareto-optimal routes, thus effectively reducing the complexity. In this section, we exploit this framework and further improve it with the aid of our EQPO algorithm. More specifically, we have relaxed the dynamic programming framework of Section~\ref{sec:mordp} for the sake of striking a better accuracy versus complexity trade-off. Additionally, we have improved the quantum-assisted process of \cite{alanis2015ndqio} for identifying the Pareto-optimal routes, so that it becomes capable of ``remembering'' the OPF identified in the previous trellis stages. We will refer to this improved quantum-assisted process as the \emph{Preinitialized-NDQIO}~(P-NDQIO) algorithm. In this context, the P-NDQIO and the EQPO algorithms are presented in Sections~\ref{subsec:preinitNDQIO} and \ref{subsec:EQPOoverview}, respectively. Let us now proceed by presenting the P-NDQIO algorithm.

\subsection{Preinitialized NDQIO algorithm}\label{subsec:preinitNDQIO}
The P-NDQIO algorithm, which is formally stated in Alg.~\ref{alg:pndqio}, is the main technique of \emph{memorization}~\cite{dasgupta2006algorithms}, thus providing a significant complexity reduction by remembering and propagating the OPF identified across the previous trellis stages to the next ones. Its memorization is performed in Step~1 of Alg.~\ref{alg:pndqio}, where the OPF of the current trellis stage is initialized to that of the previous stage. Subsequently, the P-NDQIO algorithm performs its iterations, looking for Pareto-optimal routes in Steps~2-14 of Alg.~\ref{alg:pndqio}.
\begin{algorithm}[h]
\caption{Preinitialized Non-Dominated Quantum Iterative Optimization Algorithm (P-NDQIO)} \label{alg:pndqio}
\begin{algorithmic}[1]
\STATE Initialize the OPF to $S^\text{OPF}_{(i)}\leftarrow S^\text{OPF}_{(i-1)}$.
\REPEAT
\STATE $\mathcal{T} \leftarrow 0$.
\STATE Invoke the BBHT-QSA of \cite[Alg.~1]{alanis2015ndqio} searching for routes in $S^\text{gen}_{(i)}$ that are not dominated by any of the routes of $S^\text{OPF}_{(i)}$ and output $x_s$.
\IF {$\mathbf{f}(j)\nsucc \mathbf{f}(x_s),~\forall j\in S^\text{OPF}_{(i)}$}
\REPEAT
\STATE Set $j \leftarrow x_s$.
\STATE Invoke the BBHT-QSA of \cite[Alg.~1]{alanis2015ndqio} searching for routes in $S^\text{gen}_{(i)}$ that dominate the route $j$ and output $x_s$.
\UNTIL {$\mathbf{f}(x_s)\nsucc \mathbf{f}(j)$}.
\STATE Discard the routes from $S^\text{OPF}_{(i)}$ that are dominated by the route $j$ and append it to the OPF.
\ELSE
\STATE Set $\mathcal{T} \leftarrow \mathcal{T} + 1$.
\ENDIF
\UNTIL {$\mathcal{T}=2$}
\STATE Export the $S^\text{OPF}_{(i)}$ and exit.
\end{algorithmic}
\end{algorithm}

During each iteration, which results in identifying a single Pareto-optimal route, the P-NDQIO algorithm first invokes the so-called \emph{Boyer-Brassard-Hoyer-Tapp Quantum Search Algorithm}~(BBHT-QSA)~\cite{PROP:PROP493} for the sake of identifying routes that are not dominated by any of the routes belonging to the hitherto identified OPF. We refer to this process as the \emph{Backward BBHT-QSA}~(BW-BBHT-QSA) process \cite{alanis2015ndqio}. If an invalid route-solution - i.e. a route that is indeed dominated by the OPF identified so far - is output by the BBHT-QSA, the P-NDQIO algorithm concludes that the entire OPF has been identified. However, since the BBHT-QSA exhibits a low probability of failing to identify a valid solution\footnote{We define a valid route-solution as the specific route that satisfies the condition in Step~5 of Alg.~\ref{alg:pndqio}}, the BW-BBHT-QSA step is repeated for an additional iteration in order to ensure the detection of the entire OPF, as seen in Steps~12 and 14 of Alg.~\ref{alg:pndqio}. Otherwise, should a valid route-solution {\color{blue} be} identified by the BW-BBHT-QSA step, this specific route is classified as ``potentially'' being Pareto-optimal. Consequently, the P-NDQIO algorithm invokes the so-called \emph{BBHT-QSA chain process}~\cite{alanis2014ndqo,alanis2015ndqio} in Steps~6-9 of Alg.~\ref{alg:pndqio}. Explicitly, the output of the BW-BBHT-QSA is set as the initial reference solution in Step~7 of Alg.~\ref{alg:pndqio} and a BBHT-QSA process is activated in Step~8 of Alg.~\ref{alg:pndqio}, which searches for routes that dominate the reference one. If a route that dominates the reference one is found, the reference route is updated to the BBHT-QSA output and a new BBHT-QSA process is activated. Naturally, the activation of the BBHT-QSA process is repeated until a particular route is output by the BBHT-QSA  that does not dominate the reference route, thus indicating that the reference route is Pareto-optimal. Subsequently, the Pareto-optimal routes of the set $S^\text{OPF}_{(i)}$ are checked as to whether they are dominated by the reference route, so that they are removed and the reference route is then included in $S^\text{OPF}_{(i)}$, as seen in Step~10 of Alg.~\ref{alg:pndqio}. Explicitly, this check, which is referred to as the \emph{OPF Self-Repair}~(OPF-SR) process in \cite{alanis2015ndqio}, provides the EQPO algorithm with resilience against including sub-optimal routes in the early trellis stages due to the limited number of generated routes, hence preventing their propagation to the later stages.

Both the BW-BBHT-QSA process and the BBHT-QSA chains are parts of the original NDQIO algorithm; thus, the P-NDQIO algorithm employs quantum circuits that are identical to those of the NDQIO algorithm. Therefore, the motivated readers may refer to \cite{alanis2015ndqio} for extended discussions.

\subsection{EQPO algorithm}\label{subsec:EQPOoverview}
The dynamic framework introduced in Section~\ref{sec:mordp}, albeit optimal in terms of its capability of identifying the entire OPF, it may impose an excessive complexity quantified in terms of the number of dominance comparisons required for solving the optimization problem of Eq.~\eqref{eq:ssurv_i_optim}. To elaborate further, as the number of UFs considered increases, the number of surviving routes is increased due to the differences among the UFs. This in turn leads to the proliferation of the number of routes generated per trellis stage. However, only a relatively small fraction of the surviving route-population leads eventually to generating Pareto-optimal routes in the next trellis stages. Therefore, the employment of the optimal dynamic framework presented in Section~\ref{sec:mordp} imposes a significant complexity overhead for the sake of ensuring the detection of the entire set of  Pareto-optimal routes. Consequently, a performance versus complexity trade-off has to be struck for the sake of mitigating this complexity overhead. In fact, this specific balance is struck in the context of the EQPO algorithm by jointly relying on Relaxations~\ref{rel:relaxed_generators} and \ref{rel:relaxed_generation}.

\begin{rel}\label{rel:relaxed_generators}
A route can only generate optimal routes based on Definition~\ref{def:route_generation}, if it is Pareto-optimal. This is formally formulated as follows:
\begin{equation}\label{eq:rel_surv}
 S^\text{surv}_{(i)}\triangleq S^\text{OPF}_{(i)}-S^\text{OPF}_{(i-1)}.
\end{equation}
\end{rel}
Relaxation~\ref{rel:relaxed_generators} restricts the set $ S^\text{surv}_{(i)}$ of the surviving routes at the end of the $i$-th trellis stage to the set of the newly-discovered Pareto-optimal routes at this specific trellis stage. This relaxation provides beneficial complexity reduction, since it makes the identification both of the set $S^\text{surv}_{(i)}$ of surviving routes and of the set $S^\text{OPF}_{(i)}$ of Pareto-optimal routes possible by simply solving the optimization problem of Eq.~(\ref{eq:sopf_i_optim}). Explicitly, Proposition~\ref{prop:route_optimality} does not conflict with Relaxation~\ref{rel:relaxed_generators}, since the Pareto-optimal routes are guaranteed to have Pareto-optimal sub-routes. This is justified by the fact that the sub-routes dominate their routes due to the absence of the final hop, which results in increasing all the UFs considered. Thus, since there exist no route from the SN to the DN dominating the route identified, there exist no routes dominating the respective sub-route either. However, the complexity reduction offered by Relaxation~\ref{rel:relaxed_generators} comes at the price of reduced accuracy, since sub-optimal routes having Parero-optimal sub-routes do exist, which might potentially lead to the generation of Pareto-optimal routes in the next trellis stages. This specific limitation is mitigated with the aid of Relaxation~\ref{rel:relaxed_generation}. 
\begin{rel}\label{rel:relaxed_generation}
For the sake of facilitating the identification of all Pareto-optimal routes, Definition~\ref{def:route_generation} is relaxed as follows: a specific route $x$ is said to generate another route $x_g^{(j,k)}$ by inserting the single RN $R_j$ between the $k$-th and the $(k+1)$-st nodes. 
\end{rel}
Relaxation~\ref{rel:relaxed_generation} extends the set $S^\text{gen}_{(i)}$ of generated routes, which are created by the set $S^\text{surv}_{(i-1)}$ of surviving routes of the previous trellis stage. This is realized by replacing a single direct link  established either by two RNs or by an RN and the DN with an indirect link involving an appropriate RN as an intermediate relay. Naturally, this specific modification enhances the heuristic accuracy of the EQPO algorithm, since it allows the generation of additional routes, thus acting similarly to the \emph{mutation operation} of \emph{genetic algorithms} \cite{Deb:NSGA_2}.

\begin{algorithm}[htb]
\caption{Evolutionary Quantum Pareto Optimization (EQPO) Algorithm\label{alg:eqpo}.}
\begin{algorithmic}[1]
\STATE Set $S^{\text{gen}}_{(0)} \leftarrow \{SN\rightarrow DN\}$,  $S^\text{OPF}_{(0)}\leftarrow S^{\text{gen}}_{(0)}$, $S^\text{surv}_{(0)}\leftarrow S^{\text{gen}}_{(0)}$, $i\leftarrow 0$.
\REPEAT
\STATE Set $i\leftarrow i+1$.
\STATE Generate the set of routes $S^{\text{gen}}_{(i)}$ from the set $S^\text{surv}_{(i-1)}$ based on Relaxation~\ref{rel:relaxed_generation} by appropriately inserting a single RN between two intermediate nodes.
\STATE Set $S^{\text{gen}}_{(i)}\leftarrow S^{\text{gen}}_{(i)}\cup S^\text{OPF}_{(n-1)}$. 
\STATE Invoke the P-NDQIO algorithm of Alg.~\ref{alg:pndqio} in the set $S^{\text{gen}}_{(i)}$ and initialize the identified OPF to $S ^\text{OPF}_{(n)}\leftarrow S ^\text{OPF}_{(n-1)}$.
\STATE Set $S^\text{surv}_{(i)}\leftarrow S^\text{OPF}_{(n)} - S^\text{OPF}_{(n-1)}$. 
\UNTIL{$\abs{S^\text{surv}_{(i)}}=0$ \OR $i=N_\text{nodes}-1$}
\STATE Export the OPF $S ^\text{OPF}_{(n)}$ and terminate.
\end{algorithmic}
\end{algorithm}

Let us now proceed by elaborating on the specifics of the EQPO algorithm, which is formally presented in Alg.~\ref{alg:eqpo}. To elaborate further, in Step~1 of Alg.~\ref{alg:eqpo} the EQPO algorithm initializes  the set of routes generated, the Pareto-optimal routes as well as the surviving routes to the direct route, i.e. to the route $\{SN\rightarrow DN\}$. It then proceeds with the trellis stages using Steps~2-8 of Alg.~\ref{alg:eqpo}. During each trellis stage, the set $S^{\text{gen}}_{(i)}$ of generated routes is constructed in Step 4 of Alg.~\ref{alg:eqpo} relying on Relaxation~\ref{rel:relaxed_generation}. Upon applying Relaxations~\ref{rel:relaxed_generators} and \ref{rel:relaxed_generation} in the trellis of Fig.~\ref{fig:optimal-trellis} results in the trellis of Fig.~\ref{fig:eqpo-trellis}.

This set is then concatenated with the set $S^{\text{OPF}}_{(i-1)}$ of Pareto-optimal routes identified in the previous stage. Subsequently, the P-NDQIO algorithm is invoked in Step~6 of Alg.~\ref{alg:eqpo} for the sake of identifying the set $S^{\text{OPF}}_{(i)}$ of Pareto-optimal routes from the set $S^{\text{gen}}_{(i)}$. Then, the set $S^{\text{surv}}_{(i)}$ of surviving routes is determined in Step~7 of Alg.~\ref{alg:eqpo}, relying on Relaxation~\ref{rel:relaxed_generators}.


More specifically, the steps carried out as part of the EQPO algorithm's dynamic programming framework during a single trellis stage are listed as follows:
\begin{figure*}[thb]
\begin{center}
\includegraphics[width=\linewidth]{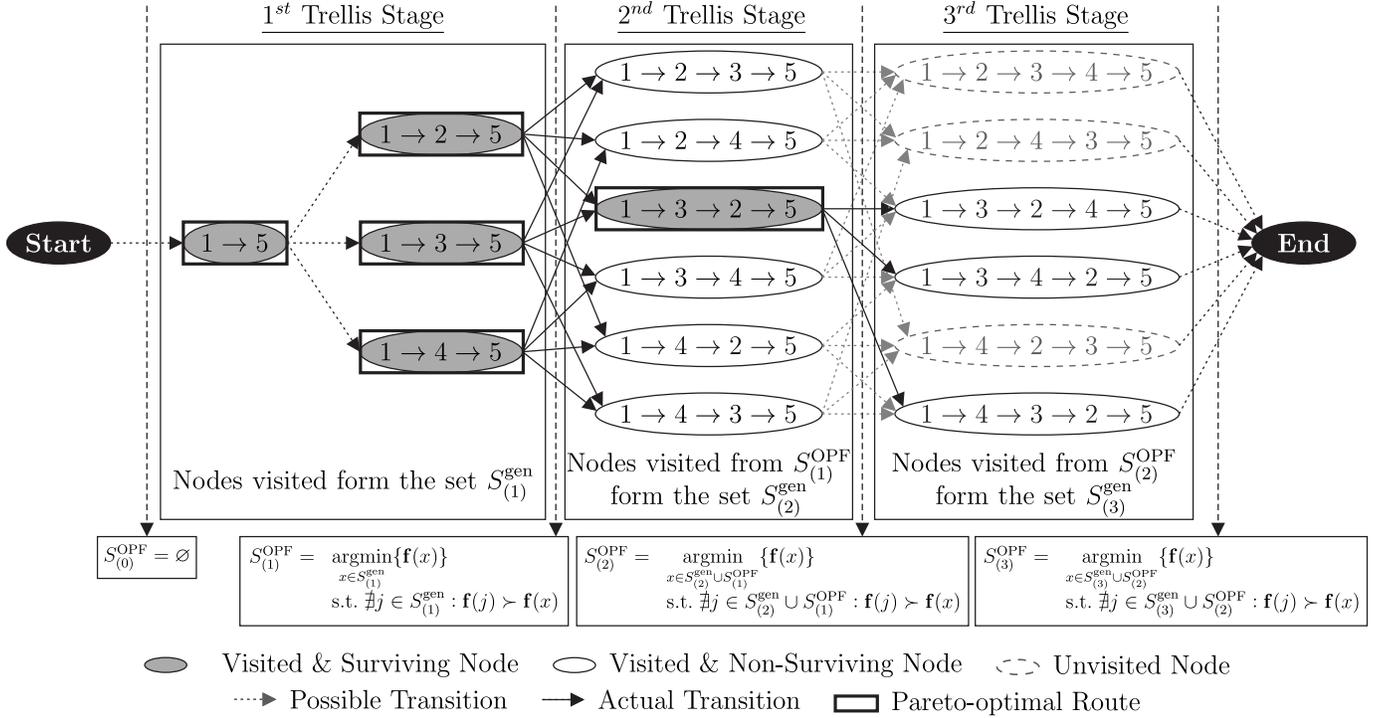}
\end{center}
\caption{Irregular trellis graph designed for guided search-space exploration for the 5-node WMHN of Fig.~\ref{fig:network-topology} using the EQPO algorithm's dynamic programming framework, encapsulated by Relaxations~\ref{rel:relaxed_generators} and \ref{rel:relaxed_generation}. Note that the UVs of each route are presented in Table~\ref{tab:tut_uvs}\label{fig:eqpo-trellis}}
\end{figure*}
\subsubsection{Route Generation} EQPO creates the set $S^\text{gen}_{(i)}$ of routes based on the set $S^\text{surv}_{(i-1)}$ of surviving routes from the previous trellis stage using Relaxation~\ref{rel:relaxed_generation}, as seen in Step~4 of Alg.~\ref{alg:eqpo}. For instance, observe in Fig.~\ref{fig:eqpo-trellis} that the route $\{1\rightarrow 2 \rightarrow 5\}$ is capable of generating 4 routes, namely the routes $\{1\rightarrow 2 \rightarrow 3 \rightarrow 5\}$, $\{1\rightarrow 2 \rightarrow 4 \rightarrow 5\}$, $\{1\rightarrow 3 \rightarrow 2 \rightarrow 5\}$, $\{1\rightarrow 4 \rightarrow 2 \rightarrow 5\}$. By contrast,  Definition~\ref{def:route_generation} allows the generation of only the first two routes, as portrayed in Fig.~\ref{fig:optimal-trellis}. Additionally, in contrast to the optimal dynamic programming framework of Section~\ref{sec:mordp}, each route of the current trellis stage in Fig.~\ref{fig:eqpo-trellis} can be generated by multiple surviving routes of the previous stage. This specific feature of Relaxation~\ref{rel:relaxed_generation} enhances the heuristic accuracy, since it enables the generation of potentially Pareto-optimal routes, which have suboptimal constructors and hence would be disregarded based on Relaxation~\ref{rel:relaxed_generators}. 
\subsubsection{Pareto-Optimal and Surviving Routes} Following the construction of the set $S^\text{gen}_{(i)}$ of the routes generated, the EQPO algorithm invokes the P-NDQIO algorithm of Section~\ref{subsec:preinitNDQIO} in Step~6 of Alg.~\ref{alg:eqpo} in order to search for new Pareto-optimal routes belonging to the set $S^\text{gen}_{(i)}$. However, based on Definition~\ref{dfn:po}, the optimality of the route depends on the set of eligible routes considered. Consequently, the OPF $S^\text{OPF}_{(i-1)}$ hitherto identified across all the previous trellis stages has to be concatenated with $S^\text{gen}_{(i)}$ in Step~5 of Alg.~\ref{alg:eqpo}, thus ensuring that the routes identified as optimal by the P-NDQIO algorithm are indeed Pareto-optimal with respect to the entire set of legitimate routes. Note that the set $S^\text{OPF}_{(i)}$ contains the Pareto-optimal routes across all trellis stages all the way up to the $i$-th one, as in the optimal dynamic programming framework of Section~\ref{sec:mordp}. Consequently, using Relaxation~\ref{rel:relaxed_generators} the Pareto-optimal routes identified at the current trellis stage are considered as surviving routes. Note that the Pareto-optimal routes identified throughout the previous stages are not taken into account, since they would generate routes already processed during the previous trellis stages. 

The EQPO algorithm continues processing the trellis stages either until it reaches a trellis stage having no surviving paths or when the maximum affordable number of trellis stages is exhausted, in a similar fashion to the optimal dynamic programming framework of Section~\ref{sec:mordp}. 

Let us now highlight the differences between the trellises of Figs.~\ref{fig:optimal-trellis} and \ref{fig:eqpo-trellis} considering the 5-node example of Fig.~\ref{fig:network-topology}. Note that the same annotation is used in Fig.~\ref{fig:eqpo-trellis} as that of Fig.~\ref{fig:optimal-trellis} Explicitly, based on Eq.~\eqref{eq:rel_surv}, the EQPO algorithm classified the specific routes, which are Pareto-optimal as being ``Pareto-Optimal'' and those that have been generated in the current stage as ``Visited \& Surviving''. Hence in contrast to Fig.~\ref{fig:optimal-trellis}, they are equivalent in Fig.~\ref{fig:eqpo-trellis}. Similar to the optimal dynamic programming framework of Section~\ref{sec:mordp}, the EQPO algorithm initializes the set $S^{\text{gen}}_{(1)}$ of generated routes to the set of the legitimate routes having either single or two hops, as portrayed in the $1^{st}$ trellis stage of Fig.~\ref{fig:eqpo-trellis}. Based on Table~\ref{tab:tut_uvs}, all the routes having two hops are Pareto-optimal and thus the EQPO algorithm classifies them as the surviving routes of the $1^{st}$ trellis stage, as seen in Fig.~\ref{fig:eqpo-trellis}. Similar to Fig.~\ref{fig:optimal-trellis}, the EQPO algorithm's trellis paths visit the entire set of routes having three hops and then the algorithm identifies the route $\{1\rightarrow 3 \rightarrow 2 \rightarrow 5\}$ as Pareto-optimal with the aid of the P-NDQIO algorithm. Consequently, this specific route is deemed to be the sole surviving route in Fig~\ref{fig:eqpo-trellis}. This is in contrast to Fig.~\ref{fig:optimal-trellis}, where three more routes have been identified as surviving ones. Recall from Fig.~\ref{fig:optimal-trellis} that these routes do not lead to Pareto-optimal routes in the last trellis stage. This in turn results in the EQPO algorithm visiting one less route in the $3^{rd}$  trellis stage, i.e. not considering the sub-optimal route $\{1\rightarrow 4 \rightarrow 2 \rightarrow 3 \rightarrow 5\}$ as potentially Pareto-optimal.

\section{Complexity versus Accuracy Discussions}\label{sec:accVScomp}
In this section, we will characterize the complexity imposed by the EQPO Alg. presented in Alg.~\ref{alg:eqpo} and evaluate its heuristic accuracy versus the complexity invested. Additionally, note that since we had no quantum computer at our disposal, the simulations of the QSAs were carried out using a classical cluster. Explicitly, since the \emph{quantum oracle gate}~$O$ \cite{nielsen2010quantum} calculates in parallel the UF vectors of all the legitimate routes in the QD, they were pre-calculated. We note that this  results in an actual complexity higher than that of the full-search method. Therefore, the employment of the quantum algorithms in a quantum computer is essential for observing a complexity reduction as a benefit of the QP. Hence, in our simulations, we have made the assumption of employing a quantum computer and we count the total number of $O$ activations for quantifying the EQPO's complexity. This number would be the same for both classical and quantum implementations. Note that in the following analysis we will use the notation $N^{x}_{(i)}\equiv\abs{S^{x}_{(i)}}$, where $N^{x}_{(i)}$ corresponds to the cardinality of the set $S^{x}_{(i)}$. 

Furthermore, our simulation results have been averaged over $10^8$ runs. During each run we have randomly generated the node's locations as well as the interference levels experienced by them with the aid of the respective distributions mentioned in Section~\ref{sec:netSec}. We have ensured that each run is uncorrelated with the rest of the runs.

Let us now proceed by analytically characterizing the complexity imposed by our proposed algorithm.

\subsection{Complexity}\label{subsec:complexity}
We will first characterize the complexity imposed by the EQPO algorithm's dynamic progamming framework, when the exhaustive search is employed instead of the P-NDQIO algorithm in Step~6 of Alg.~\ref{alg:eqpo}. We will refer to this method as the \emph{Classical Dynamic Programming}~(CDP) method and we will use it as a benchmarker for assessing the complexity reduction offered by the QP.

Prior to characterizing the EQPO algorithm and the CDP method we will analyze the the orders of the number $N_{(i)}^{\text{surv}}$ of the surviving routes and of the number $N_{(i)}^{\text{OPF}}$ of the Pareto-optimal routes identified across the first $i$ trellis stages. As far as the number $N_{(i)}^{\text{OPF}}$ of the Pareto-optimal routes identified across the first $i$ trellis stages is concerned, the trellis graph guiding the search for Pareto-optimal routes identifies more Pareto-optimal routes, as it proceeds through more trellis stages. Explicitly, its order can be formally expressed as follows: 
\begin{equation}\label{eq:NOPF_order}
O(N_{(i)}^{\text{OPF}}) = O(a_iN_{\text{OPF}})=O(N_{\text{OPF}}),~\forall i\in \{1,\dots,N_\text{nodes}-1\},
\end{equation}
where $a_i$ corresponds to the fraction of the OPF identified by the first $i$ trellis stages. Naturally, this fraction $a_i$ approaches unity as the number $i$ of trellis stages moves closer to the maximum number of hops.

Moving on to the number $N_{(i)}^{\text{surv}}$ of the surviving routes at the $i$-stage, it is equal to the number of Pareto-optimal routes identified at the $i$-th trellis stage, based on Relaxation~1. Explicitly, $N_{(i)}^{\text{surv}}$ is a fraction of the total number $N_{(i)}^{\text{OPF}}$ of the Pareto-optimal routes identified across the first $i$ trellis stages. Hence, we have $N_{(i)}^{\text{surv}} = b_iN^{\text{OPF}}_{(i)}$ with $b_i\leq 1$ $\forall i\in \{1,\dots,N_\text{nodes}-1\}$, since the set $S_{(i)}^{\text{surv}}$ of Pareto-optimal routes at the $i$-th trellis stage is included in the set $S_{(i)}^{\text{OPF}}$ of Pareto-optimal routes identified at the first $i$ trellis stages. Therefore we can evaluate the order $O(N_{(i)}^{\text{surv}})$ as follows:
\begin{equation}\label{eq:Nsurv_order}
O(N_{(i)}^{\text{surv}}) = O(b_iN^{\text{OPF}}_{(i)}) \mathop{=}\limits^{\eqref{eq:NOPF_order}} O(b_ia_iN_{\text{OPF}})= O(N_{\text{OPF}}).
\end{equation}
Consequently, in Eqs.~\eqref{eq:NOPF_order} and \eqref{eq:Nsurv_order}, we have upper bounded the order $O(N_{(i)}^{\text{surv}})$ of the number of surviving routes at the $i$-th stage as well as the order $O(N_{(i)}^{\text{OPF}}) $ of the number of  Pareto-optimal routes identified at the first $i$ stages by the order $O(N_{\text{OPF}})$ of the total number of Pareto-optimal routes, i.e. we have $ O(N_{(i)}^{\text{surv}})=O(N_{(i)}^{\text{OPF}})=O(N_{\text{OPF}})$. Naturally, Eq.~\eqref{eq:NOPF_order} and  \eqref{eq:Nsurv_order} will facilitate the complexity analysis, since they render the aforementioned orders independent of the index $i$ of the trellis stages. Let us now proceed by characterizing the complexity imposded by the CDP method.
\subsubsection{CDP method's complexity}
Let us assume that there is a total of $N^\text{gen}_{(i)}$ generated routes arriving at the $i$-th trellis stage. These particular routes are generated by the specific Pareto-optimal routes identified at the previous trellis stage, which are $N^\text{surv}_{(i-1)}$ in total. Based on the aforementioned assumptions, the number of generated routes arriving at the $i$-th trellis stage is formulated as follows: 
\begin{align}
N^\text{gen}_{(i)} &= N^\text{surv}_{(i-1)}(N_\text{nodes}-1-i)i=O\left[N^\text{surv}_{(i-1)}N_\text{nodes}i\right],\nonumber\\
& \mathop{=}\limits^{\eqref{eq:Nsurv_order}}O\left[N_\text{OPF}N_\text{nodes}i\right]\label{eq:Ngen_i}.
\end{align}
Since the set of Pareto-optimal routes of the previous trellis stage are concatenated to the set of generated routes in Step~5 of Alg.~\ref{alg:eqpo}, the total number of routes considered at the $i$-th trellis stage is given by: 
\begin{equation}\label{eq:Nroutes_i}
N^\text{routes}_{(i)} = N^\text{gen}_{(i)}+N^\text{OPF}_{(i-1)}\mathop{=}\limits^{\eqref{eq:NOPF_order},\eqref{eq:Ngen_i}}O\left[N_\text{OPF}N_\text{nodes}i\right].
\end{equation}
Additionally, the CDP method performs $O[(N^\text{routes}_{(i)})^2]$ dominance comparisons, which we will refer to as the \emph{Cost Function Evaluation}~(CFE), since each generated route has to be compared to all of the routes considered.  Therefore, the total complexity imposed by the CDP method across all trellis stages may be quantified in terms of the number of dominance comparisons, which is formulated as follows:
\begin{equation}\label{eq:Lcdp}
L_\text{CDP}=\sum\limits_{i=1}^{N_\text{nodes}-1}O\left[\left(N^\text{routes}_{(i)}\right)^2\right]= O(N_\text{OPF}^2N_\text{nodes}^5),
\end{equation} 
where we have exploited the property of the sum of squared numbers \cite{abramowitz1966handbook}, where we have $\sum^{n}_{i=1}{i^2}=O(n^3)$.

\subsubsection{EQPO algorithm's complexity}
Moving on to the EQPO algorithm's complexity analysis, the P-NDQIO algorithm is activated once per trellis stage, based on Alg.~\ref{alg:eqpo}.  Note that we will classify the complexity imposed by the P-NDQIO into two different domains, namely that of the parallel and that of the sequential complexity. To elaborate further, the P-NDQIO algorithm also exploits the synergies between QP and HP, which was utilized by the NDQIO algorithm of \cite{alanis2015ndqio}. Explicitly, the parallel complexity, which is termed as ``normalized execution time'' in \cite{alanis2015ndqio},  is defined as the number of dominance comparisons, when taking into account the degree of HP. Therefore, it may be deemed to be commensurate with the algorithm's actual normalized execution time. By contrast, the sequential complexity, which is termed as ``normalized power consumption'' in \cite{alanis2015ndqio}, is defined as the total number of dominance comparisons, without considering the potential degree of HP. Hence, this specific complexity may be deemed to be commensurate with the algorithm's normalized power consumption, as elaborated in \cite{alanis2015ndqio} as well.

Let us now proceed by characterizing the complexity of the individual sub-processes of the P-NDQIO process. During each trellis stage, the P-NDQIO algorithm activates its BW-BBHT-QSA step. This step invokes the BBHT-QSA once; however, since the quantum circuits of the original NDQIO algorithm are utilized, each activation of the quantum oracle, namely the operator $U_G$ in \cite[Fig.~8]{alanis2015ndqio}, compares each of the generated routes to all the routes comprising the OPF identified so far. Since this set of comparisons is carried out in parallel, a single activation imposes a single CFE and $N^\text{OPF}_{(i)}$ CFEs in the parallel and sequential domains, respectively. Note that the BW-BBHT-QSA process will be activated $(N^\text{surv}_{(i)}+2)$ times during a single trellis stage, since we opted for repeating this step for an additional iteration, when the BBHT-QSA fails to identify a valid route. Therefore, the parallel and sequential complexity imposed by the BW-BBHT-QSA process are quantified as follows:
\begin{align}
L^{BW,P}_{(i)} &= (N^\text{surv}_{(i)}+2)L_\text{BBHT}(N^\text{routes}_{(i)}),\label{eq:Lbw_P}\\
&=O(N_\text{OPF}\sqrt{N_\text{OPF}N_\text{nodes}i}),\label{eq:Lbw_P_ord}\\
L^{BW,S}_{(i)} &=\sum\limits_{j=0}^{N^\text{surv}_{(i)}}{(j+N^\text{OPF}_{(i-1)})}~L_\text{BBHT}(N^\text{routes}_{(i)})+\nonumber\\ &\;\;\;\; +N^\text{surv}_{(i)}L_\text{BBHT}(N^\text{routes}_{(i)})\label{eq:Lbw_S},\\
&= O(N_\text{OPF}^2\sqrt{N_\text{OPF}N_\text{nodes}i}).\label{eq:Lbw_S_ord}
\end{align}
Recall that the term $N^\text{surv}_{(i)}$ in Eqs.~\eqref{eq:Lbw_P} and \eqref{eq:Lbw_S} corresponds to the number of Pareto-optimal routes identified . Additionally, for the calculation of the orders of complexity we have relied on the fact that the BBHT-QSA has a complexity on the order of $L_\text{BBHT}(N)=O(\sqrt{N})$ as demonstrated both in \cite{alanis2015ndqio} and in \cite{PROP:PROP493}. Moving on to the complexity imposed by the BBHT-QSA chains, it has been demonstrated in \cite{alanis2015ndqio} that the complexity imposed by a single of BBHT-QSA chain - which leads to the identification of a single Pareto-optimal route - is identical to that of the so-called \emph{Durr-Hoyer Algorithm}~(DHA) \cite{durr1996quantum}, namely on the order of $L_{\text{DHA}}(N)=O(\sqrt{N})$ in terms of the number of quantum oracle gate activations. As for the latter, the $U_{g^\prime}$ quantum operator of \cite[Fig.~7]{alanis2015ndqio} has been utilized, which implements a dominance comparison. Explicitly, each activation of this operator imposes a parallel complexity of $1/K$ CFEs and a sequential complexity of a single CFE, owing to the parallel implementation of the UF comparisons. Therefore, the parallel and sequential complexity imposed by the BBHT-QSA chains are quantified as follows:
\begin{align}
L^{chain,P}_{(i)} &= \frac{N^\text{surv}_{(i)}}{K}L_\text{DHA}(N^\text{routes}_{(i)}),\label{eq:Lchain_P}\\
&=O(N_\text{OPF}\sqrt{N_\text{OPF}N_\text{nodes}i}),\label{eq:Lchain_P_ord}\\
L^{chain,S}_{(i)} &= N^\text{surv}_{(i)}L_\text{DHA}(N^\text{routes}_{(i)}),\label{eq:Lchain_S}\\
&=O(N_\text{OPF}\sqrt{N_\text{OPF}N_\text{nodes}i}).\label{eq:Lchain_S_ord}
\end{align}
Finally, as for the OPF-SR dominance comparisons of Step~10 of Alg.~\ref{alg:pndqio}, the parallel and sequential complexity imposed by this process are quantified as follows:
\begin{align}
L^{SR,P}_{(i)} &= \frac{1}{K}\sum\limits_{j=1}^{N^\text{surv}_{(i)}}(j+N^\text{OPF}_{(i-1)})=O(N_\text{OPF}^2),\label{eq:Lsr_P}\\
L^{SR,S}_{(i)} &= \sum\limits_{j=1}^{N^\text{surv}_{(i)}}(j+N^\text{OPF}_{(i-1)})=O(N_\text{OPF}^2).\label{eq:Lsr_S}
\end{align}
Recall from Eqs.~\eqref{eq:Lbw_P_ord}, \eqref{eq:Lbw_S_ord}, \eqref{eq:Lchain_P_ord}, \eqref{eq:Lchain_S_ord}, \eqref{eq:Lsr_P} and \eqref{eq:Lsr_S} that we used Eqs.~\eqref{eq:NOPF_order} and \eqref{eq:Nsurv_order}, where we have $O(N^\text{surv}_{(i)})=O(N^\text{OPF}_{(i)})=O(N_\text{OPF})$ with $N_\text{OPF}$ corresponding to the total number of Pareto-optimal routes. Let us now proceed with the evaluation of the total parallel and sequential complexities of the EQPO algorithm. In the worst-case scenario the EQPO algorithm will process $(N_\text{nodes}-1)$ trellis stages, corresponding to the maximum possible number of hops, whilst visiting each node at most once. Thus, the total parallel and sequential complexities imposed by the EQPO algorithm are quantified as follows:
\begin{align}
L_{EQPO}^{P} &= \sum\limits_{i=1}^{N_\text{nodes}-1}{L^{BW,P}_{(i)}+L^{chain,P}_{(i)}+L^{SR,P}_{(i)}},\label{eq:Leqpo_P}\\
&= O(N_\text{OPF}^{3/2}N_\text{nodes}^2),\label{eq:Leqpo_P_ord}\\
L_{EQPO}^{S} &= \sum\limits_{i=1}^{N_\text{nodes}-1}{L^{BW,S}_{(i)}+L^{chain,S}_{(i)}+L^{SR,S}_{(i)}},\label{eq:Leqpo_S}\\
&= O(N_\text{OPF}^{5/2}N_\text{nodes}^2)\label{eq:Leqpo_S_ord}.
\end{align} 
Note that in Eqs.~\eqref{eq:Leqpo_P_ord} and \eqref{eq:Leqpo_S_ord} we have exploited the specific property of the sum of square roots, where we have $\sum_{i=1}^{n}{\sqrt{i}}=O(n^{3/2})$ \cite{abramowitz1966handbook}.  Observe from Eqs.~\eqref{eq:Lcdp} and \eqref{eq:Leqpo_P_ord} that the EQPO algorithm achieves a parallel complexity reduction against the CDP method by a factor on the order of $O(N^3_\text{nodes}\sqrt{N_\text{OPF}})$. Additionally, the respective sequential complexity reduction is by a factor on the order of $O(N^3_\text{nodes}/\sqrt{N_\text{OPF}})$, based on Eqs.~\eqref{eq:Lcdp} and \eqref{eq:Leqpo_S_ord}. Hence, the EQPO imposes a lower sequential complexity than the CDP method, as long as we have $O(N_\text{nodes}^3)>O(\sqrt{N_\text{OPF}})$. As far as the EQPO algorithm's  predecessors are concerned, it has been proven in \cite{alanis2015ndqio} that the NDQO algorithm imposes identical parallel and sequential complexities, which are on the order of $O(N\sqrt{N})$. By contrast, the NDQIO algorithm imposes a parallel and a sequential complexity, which are on the order of $O(N_\text{OPF}\sqrt{N})$ and $O(N^2_\text{OPF}\sqrt{N})$, respectively, where $N$ corresponds to the total number of legitimate routes. Consequently, the complexity imposed by both the NDQO and the NDQIO algorithms is proportional to $O(\sqrt{N})$ in both domains, yielding an exponential increase in the order of complexity as the number nodes increases. By contrast, both the EQPO algorithm and the CDP method exhibit a complexity order similar to polynomial scaling, since its has been demonstrated in \cite[Fig.~11]{alanis2015ndqio} that the total number $N_{\text{OPF}}$ of Pareto-optimal routes increases at a significantly lower rate than that of the total number $N$ of routes.

\begin{figure*}[htb]
\centering
\subfloat[Parallel Complexity\label{subfig:parallel-complexity}]{\includegraphics[width=0.5\linewidth]{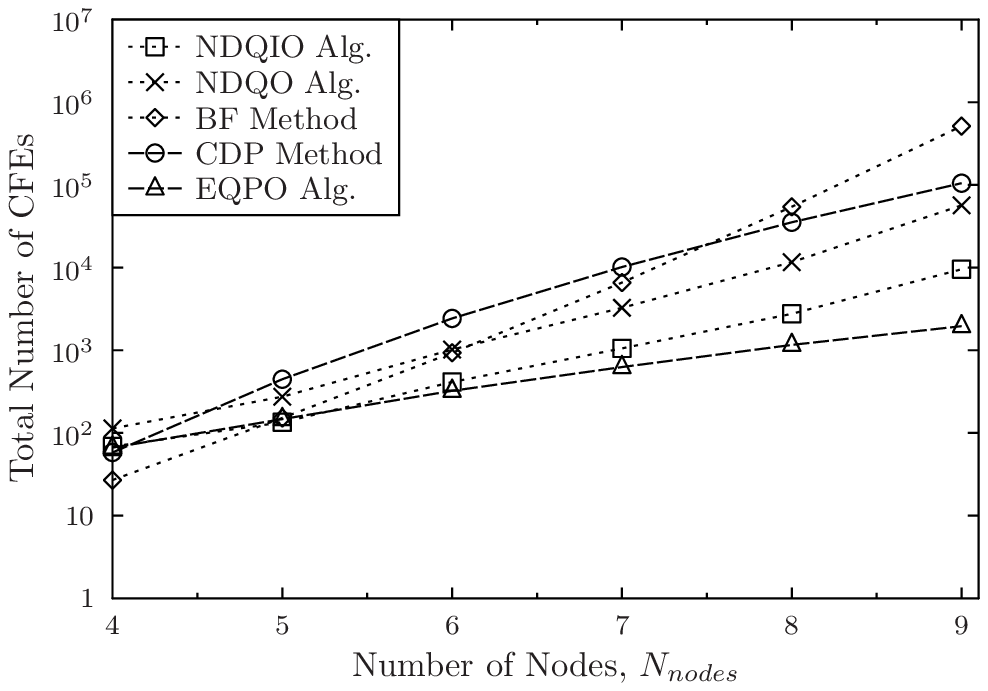}} \hfill
\subfloat[Sequential Complexity\label{subfig:sequential-complexity}]{\includegraphics[width=0.5\linewidth]{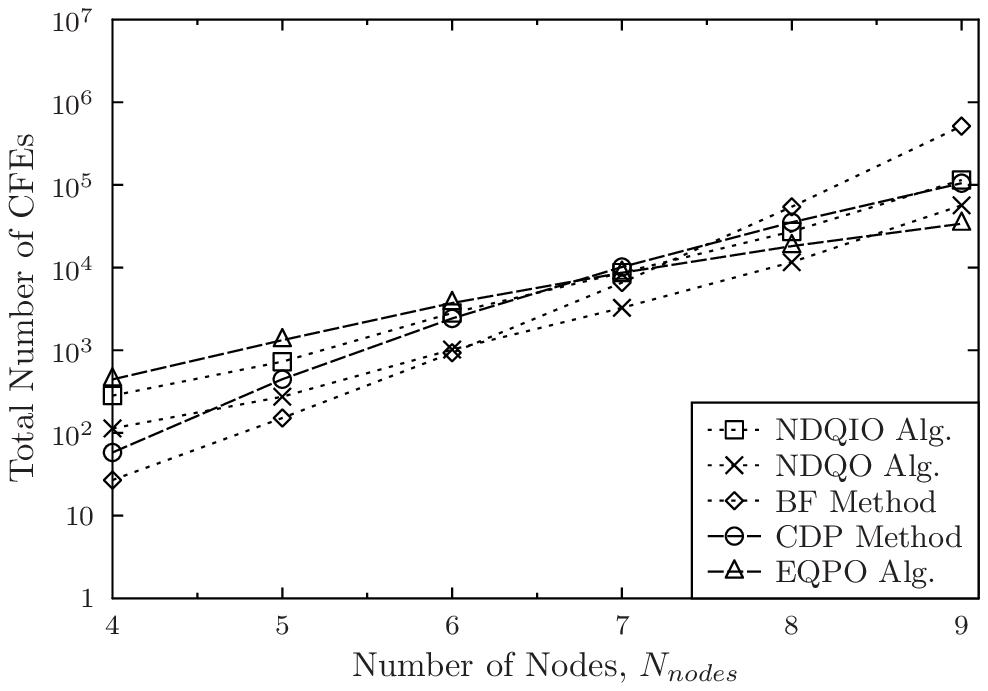}}\hfill
\caption{EQPO Alg. (a)~parallel and (b)~sequential complexity quantified in terms of the number of CFEs. The results have been averaged over $10^8$ runs.\label{fig:complexity}}
\end{figure*}
Let us now proceed by presenting the average parallel and the average sequential complexity imposed both by the EQPO algorithm and by the CDP method, which are shown in Figs.~\ref{subfig:parallel-complexity} and \ref{subfig:sequential-complexity}, respectively. We will compare the complexities imposed by the aforementioned algorithms to those of the \emph{Brute-Force}~(BF) method as well as to those of the EQPO algorithm's predecessors, namely the NDQO and the NDQIO algorithms. The aforementioned methods consider the entire set of legitimate routes, hence they have no database correlation exploitation capabilities. Additionally, the NDQO algorithm and the BF method do not employ any HP scheme, thus their respective parallel and sequential complexities are identical. As far as the average complexity of the CDP method is concerned, observe in Figs.~\ref{subfig:parallel-complexity} and \ref{subfig:sequential-complexity} that it requires a higher number of CFEs than the BF method for WMHNs having less than 8 nodes. This parallel complexity overhead is justified by the fact that the number $N_{\text{OPF}}$ of Pareto-optimal routes w.r.t. the total number $N$ of legitimate routes is relatively high. This in turn yields an increase in the fraction of trellis nodes that are classified as survivors, hence leading to more dominance comparisons. However, this trend is reversed for WMHNs having more than 7 nodes, where the CDP method exhibits a complexity reduction compared to the BF method. More specifically, for WMHNs constituted by 9 nodes, this complexity reduction is close to an order of magnitude. Still referring to 9-node WMHNs, the CDP method imposes a slightly higher parallel complexity than that of the NDQO algorithm, while it matches the sequential complexity of the NDQIO algorithm for the same 9-node network, based on Figs.~\ref{subfig:parallel-complexity} and \ref{subfig:sequential-complexity}, respectively.

Moving on to the average parallel complexity of the EQPO algorithm, observe in Fig.~\ref{subfig:parallel-complexity} that the EQPO algorithm imposes fewer CFEs than the rest of the algorithms considered for WHMNs having more than 5 nodes. Explicitly, this complexity reduction becomes more substantial, as the number of nodes increases, reaching a parallel complexity reduction of almost an order of magnitude for 9-node WMHNs, when compared to the NDQIO algorithm, which is capable of exploiting the HP as well. As for its sequential complexity, observe in Fig.~\ref{subfig:sequential-complexity} that the EQPO algorithm imposes more CFEs than the rest of the algorithms for WMHNs having less than 7 nodes. This may be justified by the relatively small number of surviving routes, which does not allow the QP to excel by providing beneficial complexity reduction. However, this trend is reversed for WMHNs having more than 6 nodes, where the number of surviving routes becomes higher. More specifically for 9-node WMHNs, the EQPO algorithm begins to impose a sequential complexity reduction w.r.t. all the remaining algorithms considered. Additionally, observe in Figs.~\ref{subfig:parallel-complexity} and \ref{subfig:sequential-complexity} that the EQPO algorithm's complexity increases with a much lower gradient, as the number of nodes increases, when compared to the full-search-based algorithms, namely to the BF method as well as to the NDQO and the NDQIO algorithms. Explicitly, this is justified by the ``almost polynomial'' order of complexity, as demonstrated in Eqs.~\eqref{eq:Leqpo_P_ord} and \eqref{eq:Leqpo_S_ord}.

\subsection{Accuracy}\label{subsec:accuracy}
Having elaborated on the complexity imposed by the EQPO let us now proceed by discussing its heuristic accuracy. Since our design target is to identify the entire set of Pareto-optimal routes, we will evaluate the EQPO algorithm's accuracy versus the complexity imposed in terms of two metrics, namely that of the average \emph{Pareto distance} $E[P_d]$ and that of the average \emph{Pareto complection} $E[C]$. The same set of metrics have been considered in \cite{alanis2015ndqio} for the evaluation of NDQIO algorithm's accuracy as well. To elaborate further, the Pareto distance of a particular route is defined as the probability of this specific route being dominated by the rest of the legitimate routes. Explicitly, given a set of Pareto-optimal routes identified by the EQPO algorithm, their average Pareto distance is a characteristic of the OPF, since it provides insights into the proximity of the exported OPF to the true OPF. Naturally, a Pareto distance having a value of $E[P_d]=0$ implies that the OPF identified by the EQPO is fully constituted by true Pareto-optimal routes. By contrast, the average Pareto completion is defined as the specific fraction of the solutions on the true OPF identified by the EQPO. Therefore, our goal is to achieve a Pareto completion as close to $E[C]=1$ as possible.

\begin{figure*}[htb]
\centering
\subfloat[\label{subfig:7-pd-par}]{\includegraphics[width=0.5\linewidth]{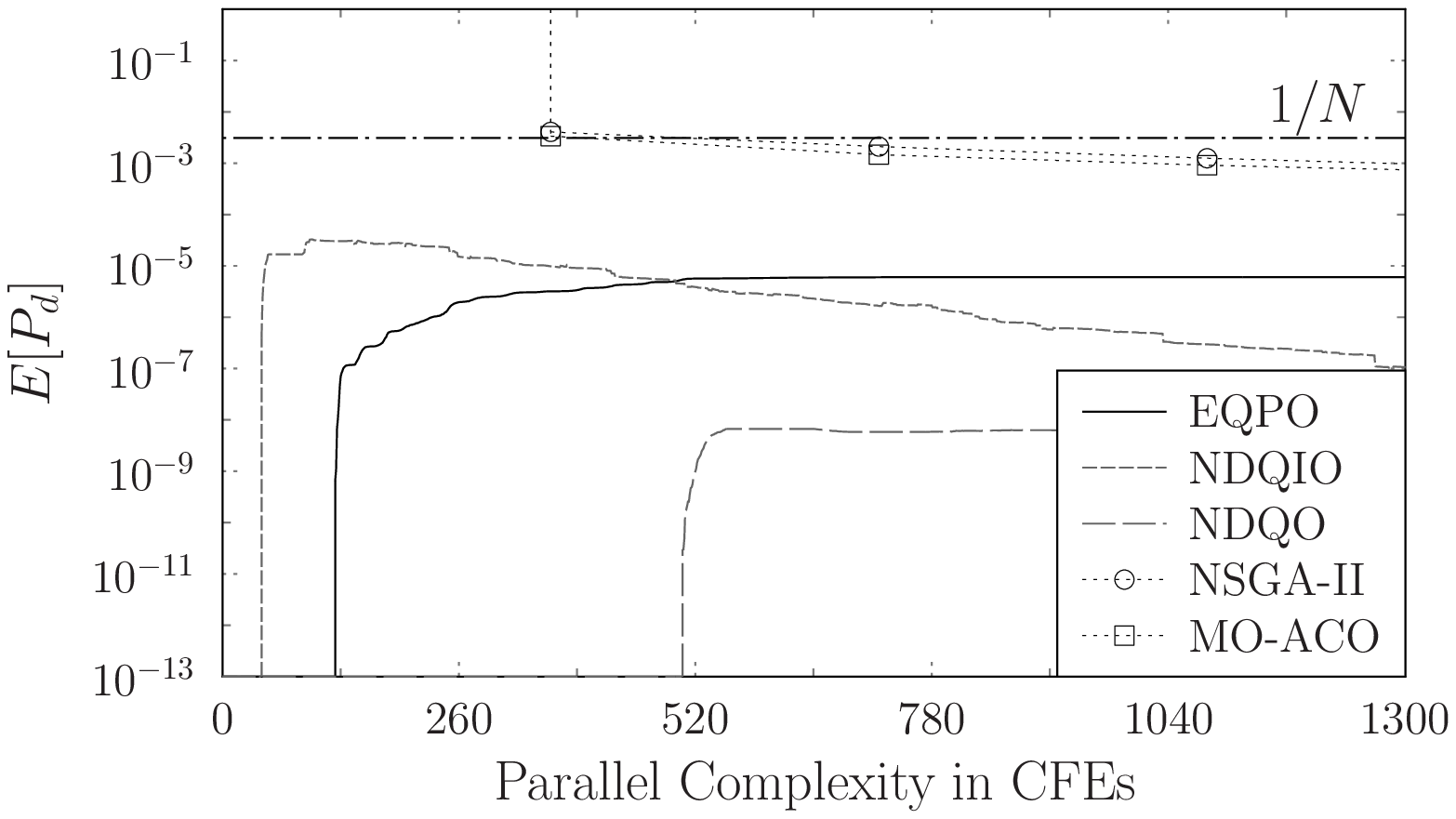}} \hfill
\subfloat[\label{subfig:7-pd-ser}]{\includegraphics[width=0.5\linewidth]{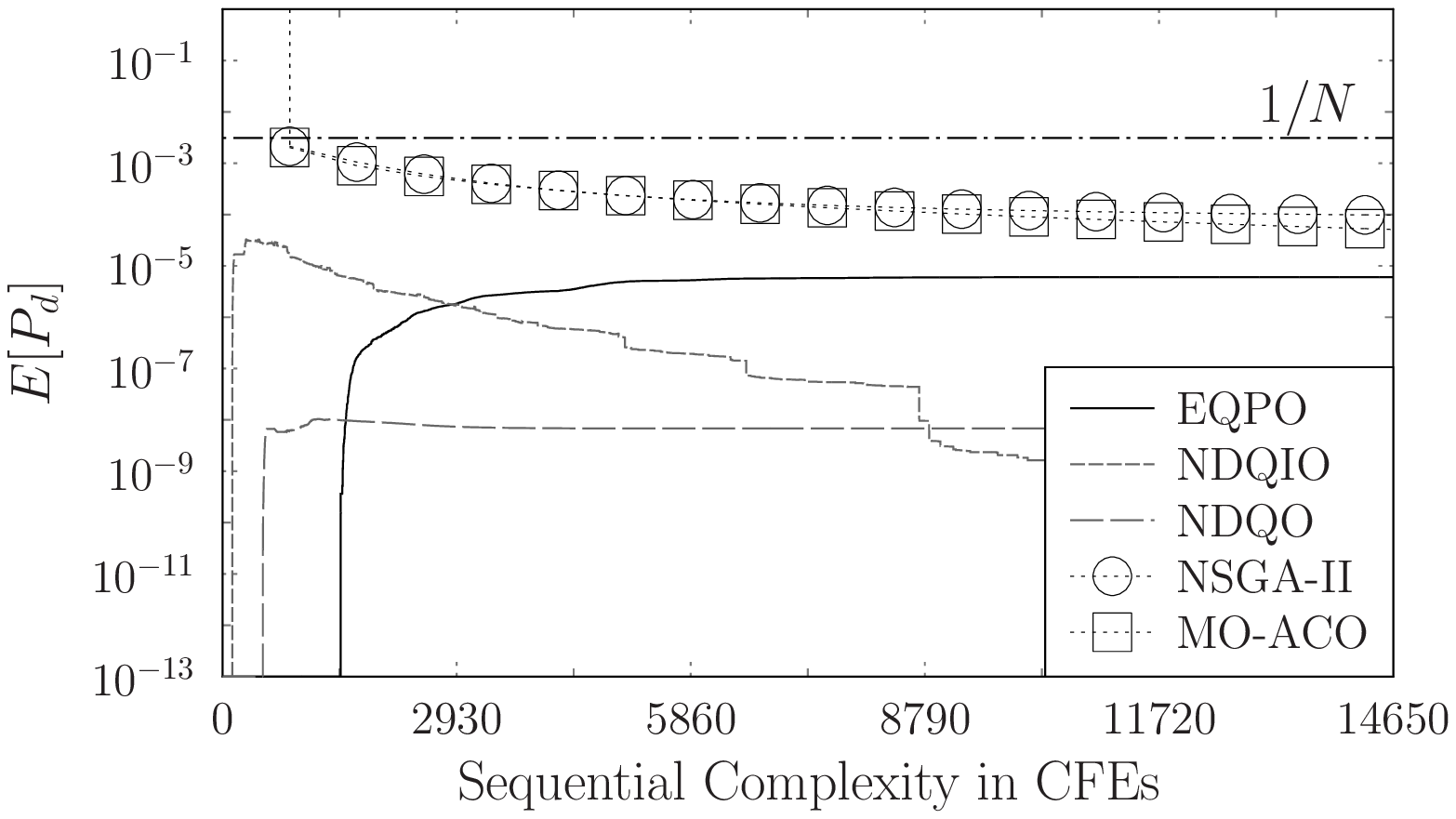}}\hfill
\subfloat[\label{subfig:7-com-par}]{\includegraphics[width=0.5\linewidth]{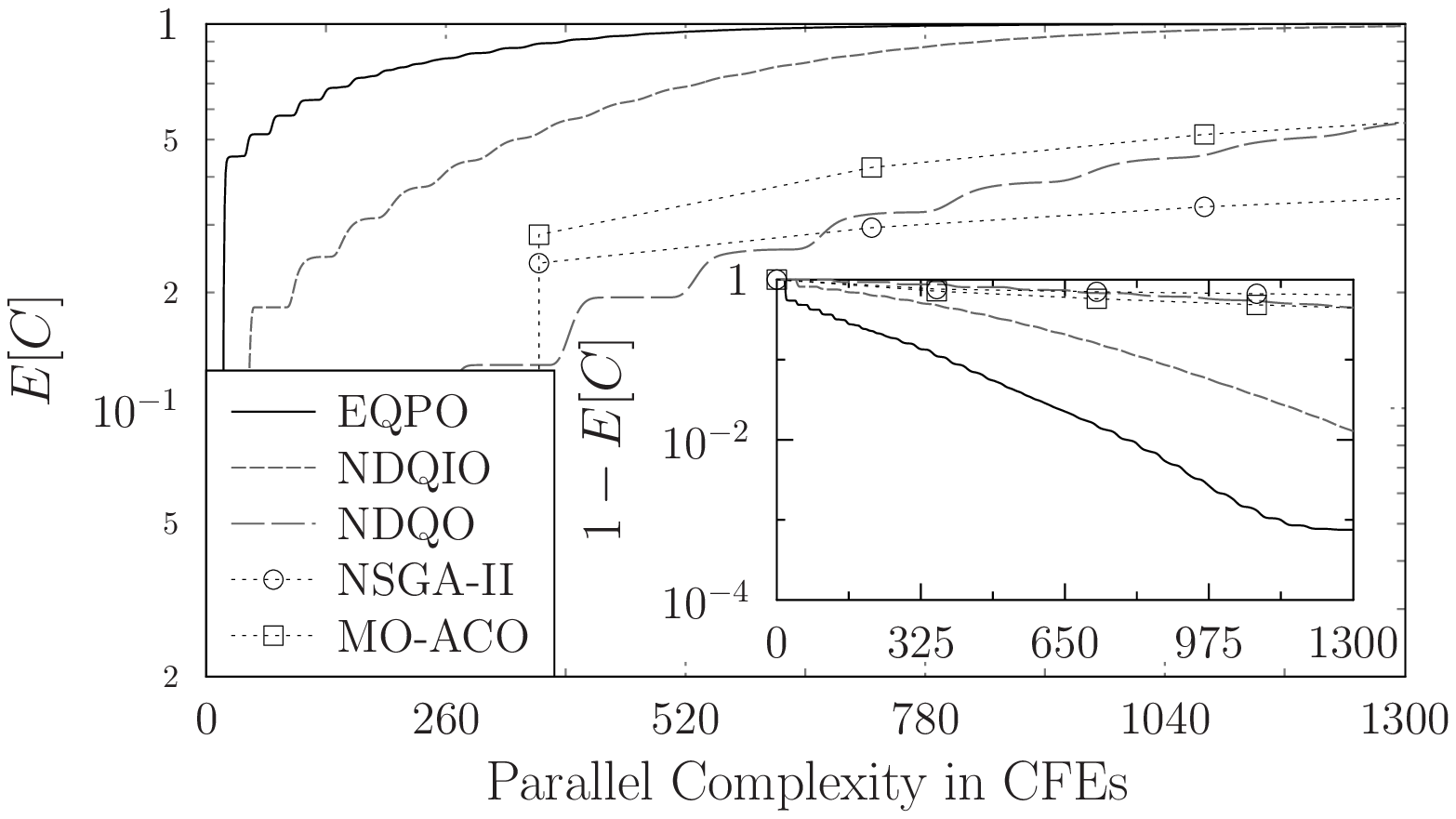}}\hfill
\subfloat[\label{subfig:7-com-ser}]{\includegraphics[width=0.5\linewidth]{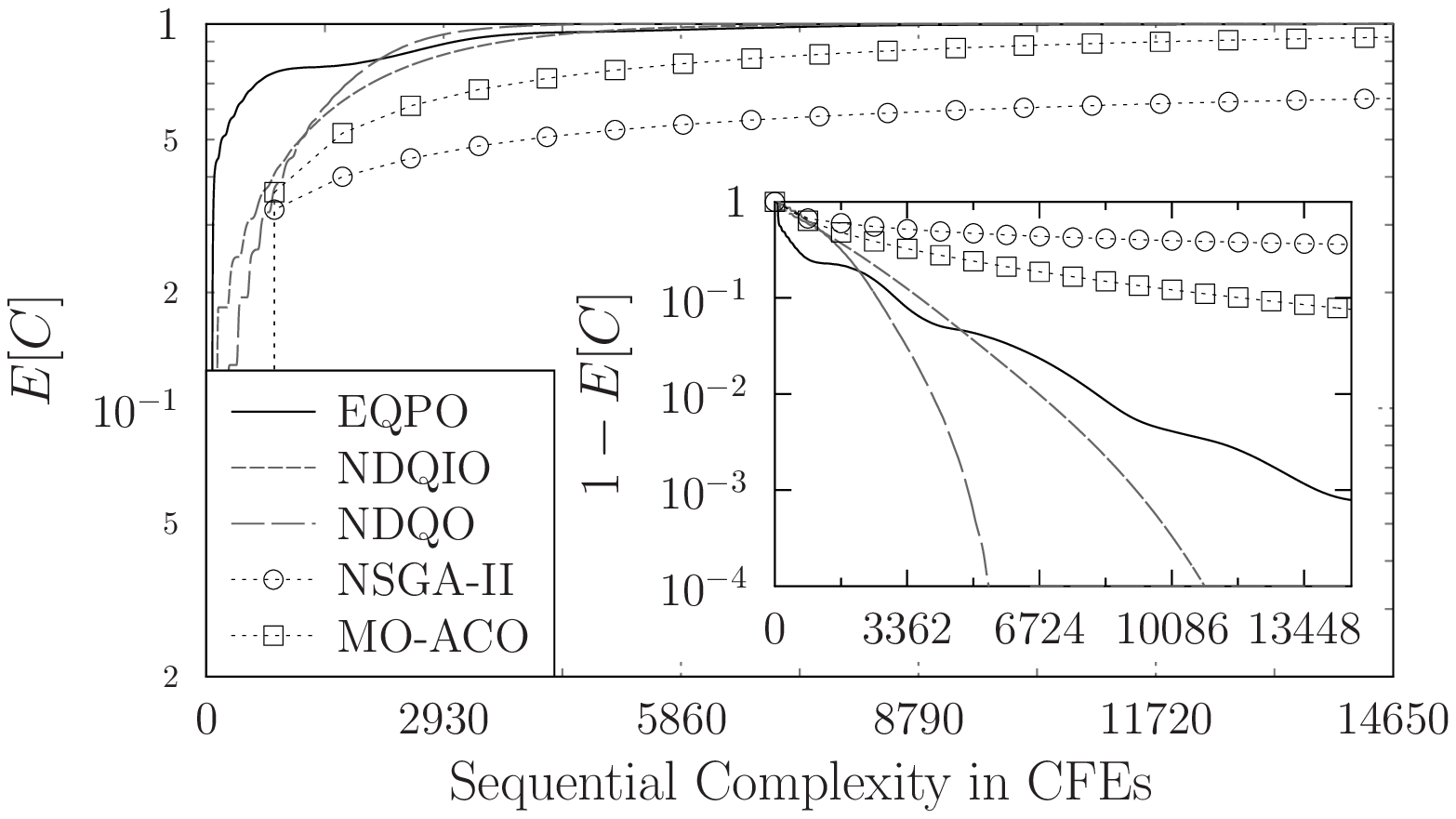}}\hfill
\caption{EQPO algorithm performance in terms of its Pareto dinstance~(a,b) and its Pareto Completion~(c,d) versus the  parallel complexity~(a,c) and the sequential complexity~(b,d) for required 7-node WHMNs. The results have been averaged over $10^8$ runs.\label{fig:accuracy-7-node}}
\end{figure*}

Having defined the performance metrics, let us now present the performance versus complexity results of the EQPO algorithm, which are shown in Fig.~\ref{fig:accuracy-7-node} for 7-node WMHNs. The reason we have evaluated the aforementioned metrics for 7-node WMHNs is for the sake of comparison to the methods analyzed in \cite{alanis2014ndqo} as well as in \cite{alanis2015ndqio}. Apart from the NDQO and NDQIO algorithms, we have used as benchmarks two additional classical evolutionary algorithms\footnote{The readers should refer to \cite{alanis2014ndqo} and to \cite{Yetgin:NSGA_2} for a detailed description of the MO-ACO and the NSGA-II, respectively.}, namely the NSGA-II and the MO-ACO. Using the same convention as in \cite{alanis2014ndqo} and \cite{alanis2015ndqio}, we have set the number of individuals equal to the number of generations and we have matched the total parallel complexity imposed by these classical algorithms to that of the NDQO algorithm, since the NDQO algorithm appears to impose the highest parallel complexity, based on Fig.~\ref{subfig:parallel-complexity}. As for their total sequential complexity we have set it to that of the NDQIO algorithm. Consequently, we have considered employing 19 individuals over 19 generations for the parallel complexity matching and 29 individuals over 29 generations for the sequential complexity matching for both the NSGA-II and the MO-ACO algorithm.

Let us now proceed by elaborating on the average Pareto distance exhibited by 7-node WMHNs versus the parallel complexity invested, as portrayed in Fig.~\ref{subfig:7-pd-par}. Observe in this figure that the EQPO algorithm performs optimally -- in the sense that no suboptimal routes are included in the OPF -- for about 130 CFEs and then exhibits an error floor around $6\cdot 10^{-6}$. Similar trends are observed for the classical NSGA-II and for MO-ACO algorithm as well as for the quantum-assisted NDQO algorithm; the classical benchmark algorithms both exhibit an error floor around $10^{-3}$, while the respective NDQO algorithm's error floor is around $7\cdot 10^{-9}$. By contrast, the NDQIO algorithm initially has an error floor of about $3\cdot 10^{-5}$, which then decays to infinitesimally low levels, when more CFEs are invested owing to its OPF-SR process \cite{alanis2015ndqio}. This specific trend is visible in Fig.~\ref{subfig:7-pd-par}, where the NDQIO algorithm outperforms the NDQO technique in terms of their $E[P_d]$ beyond 8842 CFEs in the sequential complexity domain. Additionally, the NDQIO algorithm begins to exhibit a lower $E[P_d]$ than that of the EQPO algorithm after 498 and 2932 CFEs in the parallel and sequential domains, respectively.

Let us now provide some further insights into the significance of the aforementioned error floors. Explicitly, a particular route is considered suboptimal, if there exists even just a single route dominating it, i.e. if it has a Pareto distance higher than or equal to $P^{th}_d=1/N$, where $N$ corresponds to the total number of legitimate routes. This threshold is visually portrayed with the aid of the dashed and dotted horizontal lines in Figs.~\ref{subfig:7-pd-par} and \ref{subfig:7-pd-ser}. Hence, we can normalize the results w.r.t. this threshold for exporting the probability of a specific route becoming suboptimal. Consequently, EQPO algorithm's error floor is translated into a probability of a specific route being suboptimal, which is equal to 0.2\%, while the respective probability of the NDQO algorithm is equal to $2\cdot 10^{-6}$. Additionally, the respective probabilities of the classical benchmark algorithms are about 33\% and 3.3\%, when parallel and sequential complexity are considered, respectively. Consequently, the EQPO algorithm's probability of opting for a suboptimal route may be regarded as negligible.

The evaluation of the average Pareto completion probability versus the parallel and the sequential complexity are shown in Figs.~\ref{subfig:7-com-par} and \ref{subfig:7-com-ser}. Note that the subplots inside these figures portray the portion of unidentified true Pareto-optimal routes, as encapsulated by the expression of $1-E[C]$. Explicitly, we will utilize this metric for assessing the error floor w.r.t. the $E[C]$, which may not be visible from the main plots. Additionally, note that we examined both $E[P_d]$ and $E[C]$ versus the parallel and sequential complexity imposed up to the maximum value observed by the EQPO algorithm. As far as the EQPO algorithm's average Pareto completion versus the parallel complexity is concerned, observe in Fig.~\ref{subfig:7-com-par} that the EQPO is capable of identifying a higher portion of the true OPF, when compared to the rest of the algorithms examined, while considering the same number of CFEs in the parallel complexity domain. Explicitly, the EQPO algorithm succeeds in identifying almost the entire set of Parero-optimal routes, since it is only incapable of identifying as few as 0.1\% of the entire true OPF. This error floor is reached after 1301 and 14651 CFEs in the parallel and sequential complexity domains, respectively, as it can be verified by Figs.~\ref{subfig:7-com-par} and \ref{subfig:7-com-ser}.

By contrast, this trend is not echoed in the sequential complexity domain. To elaborate further, observe in Fig.~\ref{subfig:7-pd-ser} that the EQPO algorithm remains more efficient than its classical counterparts. On the other hand, while it is indeed more efficient than the NDQO algorithm up to a complexity budget of 2147 sequential CFEs, it identifies less Pareto-optimal routes than the NDQO algorithm. The same trend is observed for the NDQIO algorithm as well for a complexity budget of 4794 sequential CFEs. Nevertheless, this discrepancy between  the parallel and sequential complexity is expected to be decreased, as the number of nodes increases. This is justified by the fact that the EQPO algorithm imposes a lower sequential complexity as the nodes proliferate, as seen in Fig.~\ref{subfig:sequential-complexity}.

Last but not least, the results portrayed on Fig.~\ref{fig:accuracy-7-node} rely on the intelligent central node having perfect knowledge both of the nodes' geo-locations and of the interference power levels experienced by them. This fundamental assumption, albeit impractical, provides us with the upper bound of the achievable performance of the routing schemes considered. Explicitly, despite its impractical nature, it facilitates a fair comparison of the EQPO algorithm to its predecessors in terms of their complexity and heuristic accuracy, which is the main focus of this treatise. Intuitively, a practical network information update process would result in both approximated and outdated network information, thus degrading the results of Fig.~\ref{fig:accuracy-7-node}, while maintaining the complexity per routing routing optimization at a similar order. Note that we plan on characterizing these imperfections and conceive a practical network information update scheme in our future work.

\section{Conclusions}\label{sec:conclusions}
In this treatise we have exploited the correlations in the formation of the Pareto-optimal routes for the sake of achieving a routing complexity reduction. In this context, we have first developed an optimal dynamic programming framework, which transforms the multi-objective routing problem into a decoding problem. However, this optimal framework imposes a high complexity. For this reason, we relaxed the aforementioned framework and proposed the EQPO algorithm, which is empowered by the P-NDQIO algorithm and thus jointly exploits the synergies between the QP and the HP along with the potential correlation in the formation of the Pareto-optimal routes. We then analytically characterized the complexity imposed by the EQPO algorithm showed that it is capable of solving the multi-objective routing problem in near-polynomial time. In fact, the EQPO achieved a parallel complexity reduction of almost an order of magnitude and a sequential complexity reduction by a factor of 3 for 9-node WMHNs. Finally, we demonstrated with the aid of simulations that this complexity reduction only imposes an almost negligible error, which was found to be 0.2\% and 0.1\% in terms of the average Pareto distance and the average Pareto completion probability for 7-node WMHNs.  

\appendices
\section{Proof of Proposition~\ref{prop:route_optimality}\label{app:proof}}
\begin{proof}
Let us consider the route $x^{(j)}_g= \{ SN {\rightarrow} \bar{R}_i{\rightarrow }R_j{\rightarrow}DN\}$ generated by the route $x$. Based on Eq.~(\ref{eq:uf_prot}), the UFs associated with this specific route are equal to:
\begin{equation}\label{eq:uf_xg}
f_k(x^{(j)}_g) = f_k(SN{\rightarrow}\bar{R}_i)+f_k(\bar{R}_i{\rightarrow}R_j)+f_k(R_j{\rightarrow}DN).
\end{equation}
Additionally, the sub-route $x^\prime$ is associated with the following UFs
\begin{equation}\label{eq:uf_xp}
f_k(x^\prime) = f_k(SN{\rightarrow}\bar{R}_i).
\end{equation}
Since we have $f_k(x)>0,\;\forall x$ from Eq.~(\ref{eq:uf_prot}), the sub-route  $x^\prime$ strongly dominates the route $x^{(j)}_g$ based on Eqs.~(\ref{eq:uf_xg}) and (\ref{eq:uf_xp}), i.e. we have $\mathbf{f}(x^\prime)\succ\mathbf{f}(x^{(j)}_g)$. Since now there is a specific route $x_d$ from the SN to DN that weakly dominates the sub-route $x^\prime$, i.e. we have $\mathbf{f}(x_d)\succeq\mathbf{f}(x^\prime)$, the route $x_d$ strongly dominates the route $x^{(j)}_g$ as well, yielding:
\begin{align}
\mathbf{f}(x_d)&\succeq\mathbf{f}(x^\prime)\succ\mathbf{f}(x^{(j)}_g)\label{eq:final_proof_1},\\
\mathbf{f}(x_d)&\succ\mathbf{f}(x^{(j)}_g)\label{eq:final_proof_2}
\end{align}
Consequently, based on Eq.~(\ref{eq:final_proof_2}) the route $x^{(j)}_g$ cannot be Pareto-optimal, since it is strongly dominated by the route $x_d$.
\end{proof}

\bibliographystyle{ieeetr} 
\bibliography{mybib}

\end{document}